\documentclass[11pt]{article}   	
\usepackage{geometry}                		
\pdfoutput=1

\usepackage{setspace}

\usepackage[square,authoryear]{natbib}
\usepackage{etoolbox}

\usepackage{graphicx}				
\usepackage[labelformat=simple]{subcaption}

\usepackage{placeins}
\usepackage[font=small,labelfont=bf]{caption}
										
\usepackage{amssymb}
\usepackage[font=small,labelfont=bf]{caption}
\usepackage{color}

\usepackage{tocloft}
\title{Enceladus's and Dione's floating ice shells\\ supported by minimum stress isostasy}
\author{Mikael Beuthe, Attilio Rivoldini, Antony Trinh\vspace{2mm}\\
\it Royal Observatory of Belgium\\
\it Avenue Circulaire 3, 1180 Brussels, Belgium\\
\it E-mail: mikael.beuthe@observatoire.be}    
\date{}							

\begin{document}

\maketitle

\begin{abstract}
Enceladus's gravity and shape have been explained in terms of a thick isostatic ice shell floating on a global ocean, in contradiction of the thin shell implied by librations.
Here we propose a new isostatic model minimizing crustal deviatoric stress, and demonstrate that gravity and shape data predict a $\rm{38\pm4\,km}$-thick ocean beneath a $\rm{23\pm4\,km}$-thick shell agreeing with -- but independent of -- libration data.
Isostatic and tidal stresses are comparable in magnitude.
South polar crust is only $7\pm4\rm\,km$ thick, facilitating the opening of water conduits and enhancing tidal dissipation through stress concentration.
Enceladus's resonant companion, Dione, is in a similar state of minimum stress isostasy.
Its gravity and shape can be explained in terms of a $\rm{99\pm23\,km}$-thick isostatic shell overlying a $\rm{65\pm30\,km}$-thick global ocean, thus providing the first clear evidence for a present-day ocean within Dione.
\end{abstract}

\vspace{\stretch{1}}

{\it \noindent Paper accepted for publication in Geophysical Research Letters\\
http://agupubs.onlinelibrary.wiley.com/agu/journal/10.1002/(ISSN)1944-8007/}

\section*{1 Introduction}

Saturn's moon Enceladus is celebrated for its huge south polar fractures venting jets of water vapor and ice particles \citep{porco2006} while its neighbor Dione is more discreet, though essential in maintaining Enceladus's eccentricity through a 2:1 orbital resonance.
The composition of Enceladus's plume shows that it originates in a subsurface ocean in contact with a silicate core \citep{postberg2011,hsu2015}, but a more detailed picture of the interior must be based on geodesy.
Inferences about the internal structure of Enceladus and Dione were first based on their long-wavelength shape \citep{thomas2007,nimmo2011} and gravity field \citep{iess2014,hemingway2016}.
These data reveal the presence of a hydrated silicate core with a radius of about three-fourths of the surface radius, but the
crust-ocean partition of the outer $\rm{}H_2O$ layer has remained controversial.
The recent measurement of Enceladus's librations \citep{thomas2016,nadezhdina2016} demonstrates that the crust is a thin ice shell floating on a global ocean.

For both satellites, the task of building interior models compatible with the gravity and shape data is complicated by strong deviations from hydrostatic equilibrium.
For synchronously rotating satellites, nonhydrostatic deviations can be estimated either from the gravity ratio $J_2/C_{22}$ or from the analog ratio for the shape (Text \ref{TextS1}).
If the satellite is hydrostatic, these shape and gravity ratios are both equal to 10/3 to first order in the flattening \citep{zharkov1985}.
Using the shape and gravity data of Cassini and taking into account second-order corrections, one concludes that the observed shape and gravity ratios of Enceladus deviate from hydrostaticity by thirty and ten percent, respectively, whereas the corresponding deviations for Dione are twice as large (Text \ref{TextS1}).
Such differences between the nonhydrostatic components of shape and gravity are indicative of isostasy, in which surface topography is mechanically supported and gravitationally compensated by a subsurface mass anomaly in such a way that below a certain depth, called the compensation depth, pressure is everywhere hydrostatic \citep{lambeck1980,lambeck1988}.
Enceladus's degree-three zonal gravity harmonic ($J_3$) also points to isostasy: it is fully nonhydrostatic and nonzero within its $3\sigma$ error interval, but only a third of what is expected from the corresponding shape harmonic.
Degree-three compensation is attributed to isostatic support of the south polar depression \citep{iess2014}.

In this paper, we reexamine the two main aspects of the method of \citet{iess2014} and \citet{mckinnon2015}: the second-order figure of equilibrium and the isostatic compensation model.
Our computation of the figure of equilibrium is more rigorous but does not make a significant numerical difference.
By contrast, our new theory of isostasy has important implications for the structure of the $\rm{}H_2O$ layer.
We put our conclusions on a firm footing by doing a Bayesian inversion of the gravity data taking into account the uncertainties on the shape and gravity.

\section*{2 The Problem with Classical Isostasy}

\begin{figure}[h]
   \includegraphics[width=14cm]{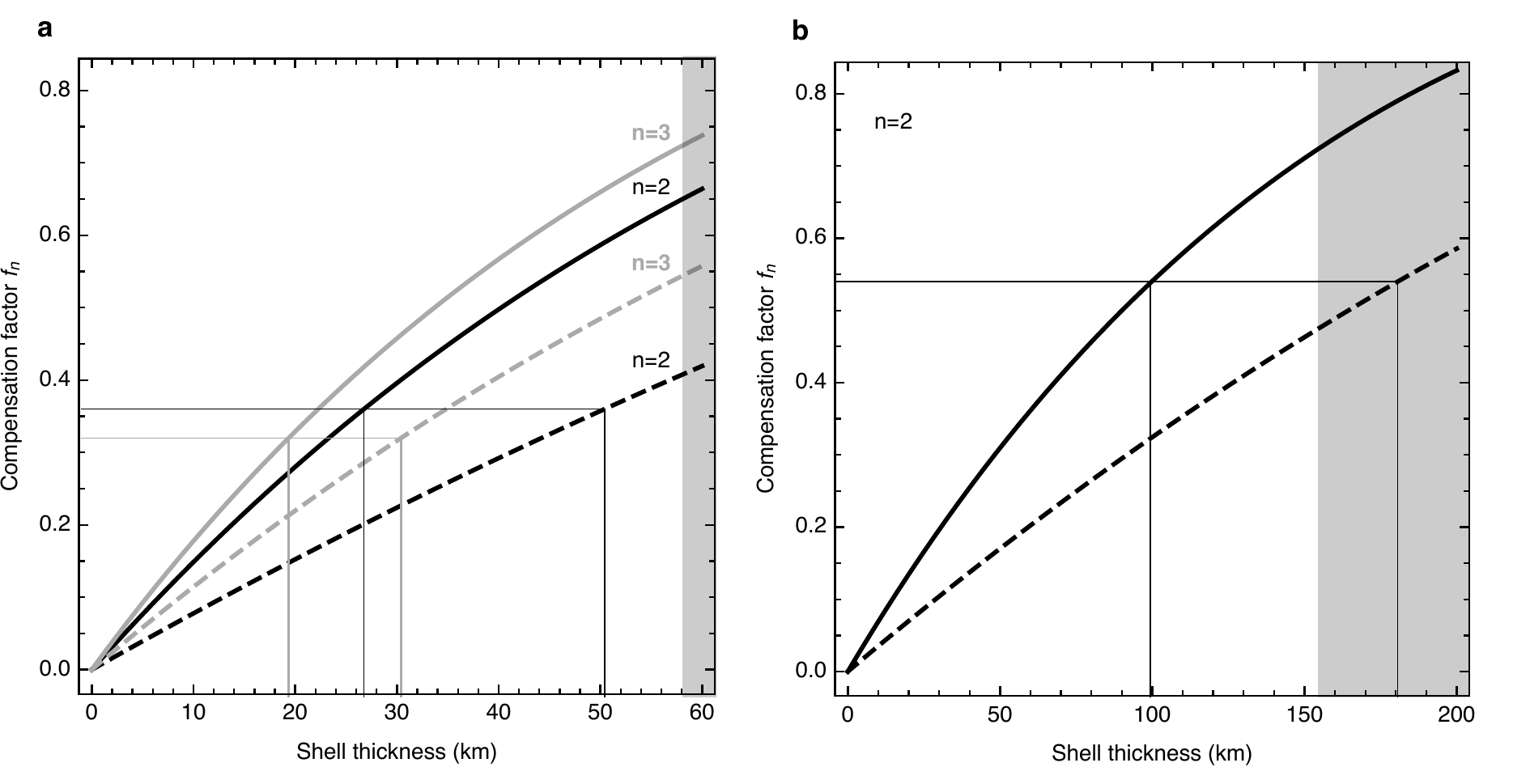}
   \caption[]{
   Compensation factor as a function of shell thickness: (a) Enceladus, (b) Dione.
   Minimum stress isostasy (solid curves) is computed for a three-layer model with an elastic core (radius of $194\rm\,km$ for Enceladus, $407\rm\,km$ for Dione), saline ocean (density $1020\rm\,kg/m^3$), and elastic shell (density $925\rm\,kg/m^3$).
   Classical Airy isostasy (dashed curves) is given by $f_n=1-(d_s/R)^n$.
    Straight lines indicate the shell thickness corresponding to (a) $f_2=0.36$, $f_3=0.32$ and (b) $f_2=0.54$.
   Shaded areas indicate crust-core overlap.
}
   \label{Fig1}
\end{figure}

Isostasy can occur either through crustal density variations (Pratt) or through variations in crustal thickness (Airy).
For Pratt isostasy, the only plausible scenario involves porosity variations close to the surface, but compensation is too high \citep{mckinnon2015}.
With Airy isostasy, Enceladus's ice shell was initially estimated to be on average 30 to $40\rm\,km$ thick \citep{iess2014}, but this value was revised to $50\rm\,km$ due to second-order tidal-rotational effects \citep{mckinnon2015}.
A thick shell is hardly compatible with the south polar activity, raises the issue of shell-core contact at the equator precluding isostasy, and does not match degree-three compensation \citep{mckinnon2015}.
Furthermore, the result conflicts with libration models predicting that the shell is half as thick \citep{thomas2016,vanhoolst2016}.
Compared to Enceladus, Dione's case is even more serious because the $180\rm\,km$-thick ice shell implied by Airy isostasy does not fit into the $150\rm\,km$-thick $\rm{}H_2O$ layer \citep{hemingway2016} (see also Fig.~\ref{Fig1}).
Although a solution is possible within $1\sigma$ uncertainties, it entails that Dione's shape deviates even more from hydrostaticity.

The easy way out is to suppose that an elastic lithosphere partly supports the load (flexural isostasy) so that the subsurface mass anomaly can be smaller and located closer to the surface.
Flexural support of long-wavelength loads, however, generates stresses that are not only larger than the tensional strength of intact ice (1 to $2\rm\,MPa$ \citep{schulson2009}) but also much above the Coulomb failure criterion for a pervasively cracked thin lithosphere \citep{mckinnon2013}.
\citet{cadek2016} nevertheless argue for partial support of Enceladus's topography by a thin lithosphere, without addressing the question of lithospheric failure.
Besides, they overestimate the elastic thickness required for top loading by a factor of ten \citep{turcotte1981,mcgovern2002,kalousova2012}.
We give a detailed rebuttal of \citet{cadek2016} in the Supporting Information (Text~\ref{TextS7}).

Classical Airy (or Pratt) isostasy is plagued by ambiguities about the isostatic prescription, i.e.\ the constraint relating the subsurface mass anomaly to the surface load \citep{vening1946}.
At the longest wavelengths, various isostatic prescriptions lead to geoid anomalies differing by more than a factor of two \citep{dahlen1982}.
This question has never been settled because of the complexity of large-scale isostasy on Earth: thermal (Pratt) and Airy isostasy are dominant in oceanic and continental crusts, respectively, while the long-wavelength geoid is explained by mantle convection.
In planetology, isostasy invariably resorts to the equal-mass prescription applied to conical columns \citep{lambeck1988}.
This prescription however neither takes into account horizontal stresses nor geoid perturbations due to the loads themselves.
Here we consider instead (following \citet{jeffreys1970} and \citet{dahlen1982}) that the only physically meaningful isostatic prescription consists in minimizing crustal deviatoric stress in a self-consistent elastic-gravitational theory.

\section*{3 Methods}

\subsection*{3.1 Hydrostatic-Isostatic Decomposition}

Before explaining the three methods particular to this paper, we recall the principle of the hydrostatic-isostatic decomposition.
Let $H_{nm}$ and $C_{nm}$ denote the cosine real harmonic coefficients of degree $n$ and order $m$ of the shape and nondimensional gravity potential, respectively.
Following \citet{iess2014}, we split the harmonic components into hydrostatic and isostatic (nonhydrostatic) components:
\begin{eqnarray}
H_{nm} &=& H_{nm}^h + H_{nm}^{iso} \, ,
\label{decompH} \\
C_{nm} &=& C_{nm}^h + C_{nm}^{iso} \, .
\label{decompC}
\end{eqnarray}
The hydrostatic components with $n=2$ and $m=(0,2)$ are determined by computing the figure of equilibrium (the other hydrostatic components vanish).
At each harmonic degree $n$, the isostatic components are related by the isotropic admittance $Z_n$ (or equivalently by the nondimensional compensation factor $f_n$) characterizing the isostatic model:
\begin{equation}
C^{iso}_{nm} = Z_n \, H^{iso}_{nm} =  \frac{3\rho_s}{(2n+1)R \bar\rho} \, f_n \, H^{iso}_{nm} \, ,
\label{admi}
\end{equation}
where $\rho_s$ is the ice shell density, $\bar\rho$ is the bulk density and $R$ is the surface radius.

\subsection*{3.2 Figure of Equilibrium}

Enceladus's rapid rotation leads to significant deviations from the first order figure of equilibrium.
Contrary to \citet{mckinnon2015}, we do not compute the figure of equilibrium with the method of \citet{tricarico2014}, which is based on nested ellipsoids, because an ellipsoidal stratification is forbidden for heterogenous hydrostatic bodies (Hamy-Pizzetti theorem, see \citet{moritz1990}).
We work instead with the classical theory of equilibrium figures extended from first order (Clairaut's equation) to second order in the flattening \citep{zharkov2004}.

Solving the second-order equations for a multilayer body yields complicated expressions for the degree-two shape and gravity coefficients (the latter normalized with respect to the mean radius).
These complete solutions to second order are closely approximated by compact formulas depending on the fluid Love numbers $k_2^F$ and $h_2^F=1+k_2^F$:
\begin{eqnarray}
\frac{1}{R}\left( H_{20}^h , H_{22}^h \right) &=& \left( - \frac{5}{6} \, (1+\epsilon_0)  , \frac{1}{4} \, (1+\epsilon_2) \right) h_2^F q \, ,
\label{HstratM} \\
\left( C_{20}^h , C_{22}^h \right) &=& \left( - \frac{5}{6} \, (1+\delta_0)  , \frac{1}{4} \, (1+\delta_2) \right) k_2^F q \, ,
\label{CstratM}
\end{eqnarray}
where $(\epsilon_0,\epsilon_2,\delta_0,\delta_2)=(76/5,44,16,64)h_2^F{}q/21$ are the second-order corrections and $q=\omega^2R^3/GM$ is the nondimensional rotational parameter (Table \ref{TableGeneral}).
Fluid Love numbers of two- and three-layer bodies are given in \citet{zharkov2004} and \citet{zharkov2010}, respectively.
We computed $C_{2m}^h$ from $H_{2m}^h$ using the second-order relations between shape and gravity (Eq.~(49) of \citet{zharkov2010}).
Our formalism for the harmonic coefficients of degree two is equivalent to the one of \citet{zharkov2010}, except that their Eqs.~(51)-(52) and (57)-(58) are wrong and must be replaced by the two equations above.

The error on Eqs.~(\ref{HstratM})-(\ref{CstratM}) is similar to the error on the method of \citet{tricarico2014}.
It is of third order in the flattening if the satellite is homogeneous.
If not, the error is formally of second order, but numerically of third order.
Compared to the method of \citet{tricarico2014}, Eqs.~(\ref{HstratM})-(\ref{CstratM}) are simpler to use and more easily applicable to multilayer bodies.
For the Bayesian inversion, we do not use Eqs.~(\ref{HstratM})-(\ref{CstratM}) but the complete solutions of the classical approach to second order (i.e.\ the error is strictly of third order).

\subsection*{3.3 Minimum Stress Isostasy}

The principle of minimum stress isostasy is based on the idea that, over time, the crust has reached the state of minimum deviatoric stress compatible with the observed topography through internal deformation, lithospheric failure and viscoelastic relaxation \citep{dahlen1981}.
The last mechanism depends on the possible viscoelastic processes and the loading timescale.
These questions cannot be answered without knowing more about the crust (rheology and thermal state) and about the origin of the putative shell thickness variations.
If desired, one can relax the assumption of minimum stress so as to include less bottom loading and more flexural support (flexural isostasy), but the model is not predictive if we do not know the elastic thickness of the lithosphere.
Our results will show that such additional parameter is not needed, and we tend to favor the simplest model that can successfully explain the observations.
We thus assume here that the crust has reached the state of minimum stress isostasy at the largest scale.

In order to minimize crustal stresses, we need to know the deformations, gravity field perturbations, and stresses due to loads acting on the top and bottom of the crust.
In terms of standard geophysical techniques, this means computing the elastic Love numbers of a self-gravitating spherically symmetric body submitted on the one hand to a surface load and, on the other, to an internal load located at the crust-ocean boundary.
The two solutions are linearly combined with an arbitrary loading ratio, which is then fixed by minimizing the second invariant of the deviatoric stress tensor.
The result takes the form of Eq.~(\ref{admi}).

We solve the elastic-gravitational problem with the incompressible propagator matrix method \citep{sabadini2004}.
The surface and internal loads are modeled as thin layers with densities (per unit surface) $\sigma_{nm}^L$ and $\sigma_{nm}^I$, respectively.
The surface load Love numbers $(k_n^L,h_n^L)$ and the internal load Love numbers $(k_n^I,h_n^I)$ are obtained as boundary values of the full solution.
By definition, the perturbation of the gravity potential $\tilde C^{iso}_{nm}$ (dimensional here: $\tilde C^{iso}_{nm}=gRC^{iso}_{nm}$) and the radial displacement of the surface $u_{nm}$ read \citep{grefflefftz2010}
\begin{eqnarray}
\tilde C^{iso}_{nm} &=& \left( 1 + k_n^L \right) U_{nm}^L + \left( x^{n+1} + k_n^I \right) U_{nm}^I \, ,
\nonumber \\
u_{nm} &=& \left( h_n^L \, U_{nm}^L+ h_n^I  \, U_{nm}^I \right)/g \, ,
\label{un}
\end{eqnarray}
where $U_{nm}^L=4\pi{}GR\sigma_{nm}^L/(2n+1)$ and $U_{nm}^I=4\pi{}GRx\sigma_{nm}^I/(2n+1)$ are the loading potentials ($g$ is the surface gravity and $x=1-d_s/R$ where $d_s$ is the shell thickness).
The former potential depends on the topography through $U_{nm}^L=g\xi_n(H_{nm}^{iso}-u_{nm})$ where $\xi_n=3\rho_s/((2n+1)\bar\rho)$, while the latter can initially  be written as $U_{nm}^I=\zeta_{n}U_{nm}^L$, where the \textit{loading ratio} $\zeta_n$ is a negative number of order unity to be fixed by the isostatic prescription.
The compensation factor follows from Eq.~(\ref{admi}):
\begin{equation}
f_n = \frac{1 + k_n^L + \zeta_n \left( x^{n+1} + k_n^I \right)}{1 + \xi_n \left( h_n^L + \zeta_n \, h_n^I \right)} \, .
\label{fn}
\end{equation}

The final step consists in computing the loading ratio with the chosen isostatic prescription.
The magnitude of deviatoric stresses can be measured with the second invariant of the deviatoric stress tensor $\tau_{\rm II}$ \citep{dahlen1982} which is proportional to the shear (or distortional) strain energy density \citep{jaeger2007}.
For each harmonic degree $n$, we minimize the total shear energy of the crust ${\cal E}_\mu$, which depends quadratically on crustal deformations (Eq.~(8.128) of \citet{dahlen1999} restricted to the crustal volume).
Since crustal deformations are linear combinations of the deformations due to the surface and internal loads, the total shear energy of the crust can be decomposed as ${\cal E}_\mu={\cal E}_\mu^{LL}+2\zeta_n{\cal E}_\mu^{LI}+\zeta_n^2{\cal E}_\mu^{II}$ where the terms $({\cal E}_\mu^{LL},{\cal E}_\mu^{LI},{\cal E}_\mu^{II})$ do not depend on $\zeta_n$.
Crustal stresses are minimum for the loading ratio $\zeta_n=-{\cal E}_\mu^{LI}/{\cal E}_\mu^{II}$, which is a computable quantity once the elastic-gravitational problem has been solved for surface and internal loads.

In minimum stress isostasy, crustal deformation is very small: $u_{nm}=h_nU_{nm}^L/g$ with $h_n=h_n^L+\zeta_nh_n^I\sim{\cal O}(10^{-3})$.
If deflection is neglected, the topography $H_{nm}^{int}$ of the crust-ocean boundary is given by
\begin{equation}
H_{nm}^{int} = \frac{\zeta_n}{x} \, \frac{\rho_s}{\rho_o-\rho_s} \, H_{nm}^{iso} \, .
\end{equation}
This extension to finite amplitude topography introduces negligible errors in the gravitational potential (Text \ref{TextS6}).
The local shell thickness is determined from the average shell thickness $d_s$ and the coefficients $H_{nm}^{iso}-H_{nm}^{int}$ (Eq.~(S.9) in Text \ref{TextS5}).

\subsection*{3.4 Bayesian Inversion}

The figure of equilibrium and the isostatic model define the forward problem: given the interior structure and the isostatic load ($H_{nm}^{iso}$), we can predict the shape and gravity coefficients.
What interests us more is the inverse problem which is nonlinear, under- or overconstrained depending on the parameter, and based on uncertain gravity and shape data.
It is thus well suited to a Bayesian inference method \citep{sambridge2011}.
The result of the Bayesian inversion is the posterior probability density function, i.e.\ the conditional probability for the parameters given the data \citep{tarantola2005,gregory2005}.
We simulate the posterior probability density function with a Metropolis-Hastings sampler.
From the generated samples, we compute for each parameter the probability density function, the mean value and the Bayesian confidence intervals (Text \ref{TextS4}).

Enceladus and Dione are modeled as three-layer incompressible bodies made of an elastic core, an ocean, and an elastic shell.
The model parameters are the densities ($\rho_s$, $\rho_o$) and thicknesses ($d_s$, $d_o$) of the ice shell and ocean (the radius $r_c$ and density $\rho_c$ of the core are derived parameters).
The prior information is described by uniform probability density functions subjected to the constraint that the shell thickness at the south pole is not negative (Text \ref{TextS4} and Fig.~\ref{FigHistoSOL1}).
For Enceladus, the uncorrelated prior ranges are $1-60\rm\,km$ for $d_s$ and $5-60\rm\,km$ for $d_o$.
For Dione, the prior ranges are $1-200\rm\,km$ for $d_s$ and $5-200\rm\,km$ for $d_o$.
Enceladus's ocean is probably similar in salinity to the ice grains in the plume ($0.5-2\%$ salt by mass \citep{postberg2011}) while the detection of silica nanoparticles sets a 4\% upper bound on the salinity \citep{hsu2015}.
The prior range for $\rho_o$ is $1000-1040\rm\,kg/m^3$ ($0-5\%$ salt by mass) and we assume the same for Dione.
Given the low pressure within the crust, the density of pure ice varies between $920$ and $930\rm\,kg/m^3$ depending on the temperature, but porous ice is lighter whereas salt-rich ice is denser.
For the three-layer body, the prior range for $\rho_s$ extends from $880\rm\,kg/m^3$ (10\% porosity in the upper half of the crust) to $960\rm\,kg/m^3$ (no porosity, similar salt content as the ocean maximum).
For Enceladus, we model a very porous crust with a four-layer model in which the crust is made of a bottom layer with fixed density ($920\rm\,kg/m^3$) and a $10\rm\,km$-thick upper layer with a prior density range of $700-920\rm\,kg/m^3$ \citep{besserer2013}.

\section*{4 Results}

Given the compensation factor, minimum stress isostasy requires a shell nearly half as thick as in classical isostasy (Fig.~\ref{Fig1}).
The compensation factor is most sensitive to the shell thickness which plays the role of compensation depth.
The shear moduli of the shell and core (radius $r_c$) are set to $\mu_s=3.5\rm\,GPa$ (pure ice) and $\mu_c=40\rm\,GPa$ (hydrated silicates), respectively.
The compensation factor is nearly independent of $\mu_s$, $\mu_c$, and $r_c$ as long as the shell and core are not fluid-like (i.e.\ $\mu_s,\mu_c\gg\bar\rho{}gR=0.05\rm\,GPa$ for Enceladus).
As a simple example, we solve Eqs.~(\ref{decompH})--(\ref{CstratM}) for the two parameters $(k_2^F,f_2)$ and the eight hydrostatic/isostatic components of degree two, assuming zero data uncertainty and an ice-dominated $\rm{}H_2O$ layer ($\rho_s=925\rm\,kg/m^3$ as in \citet{mckinnon2015}).
For Enceladus, $(k_2^F,f_2)=(0.93,0.36)$ and $r_c=194\rm\,km$, while $(k_2^F,f_2)=(0.92,0.54)$ and $r_c=407\rm\,km$ for Dione.
Fig.~\ref{Fig1} shows that classical Airy isostasy predicts a very thin ocean for Enceladus and no ocean at all for Dione.

\begin{figure}
   \includegraphics[width=14cm]{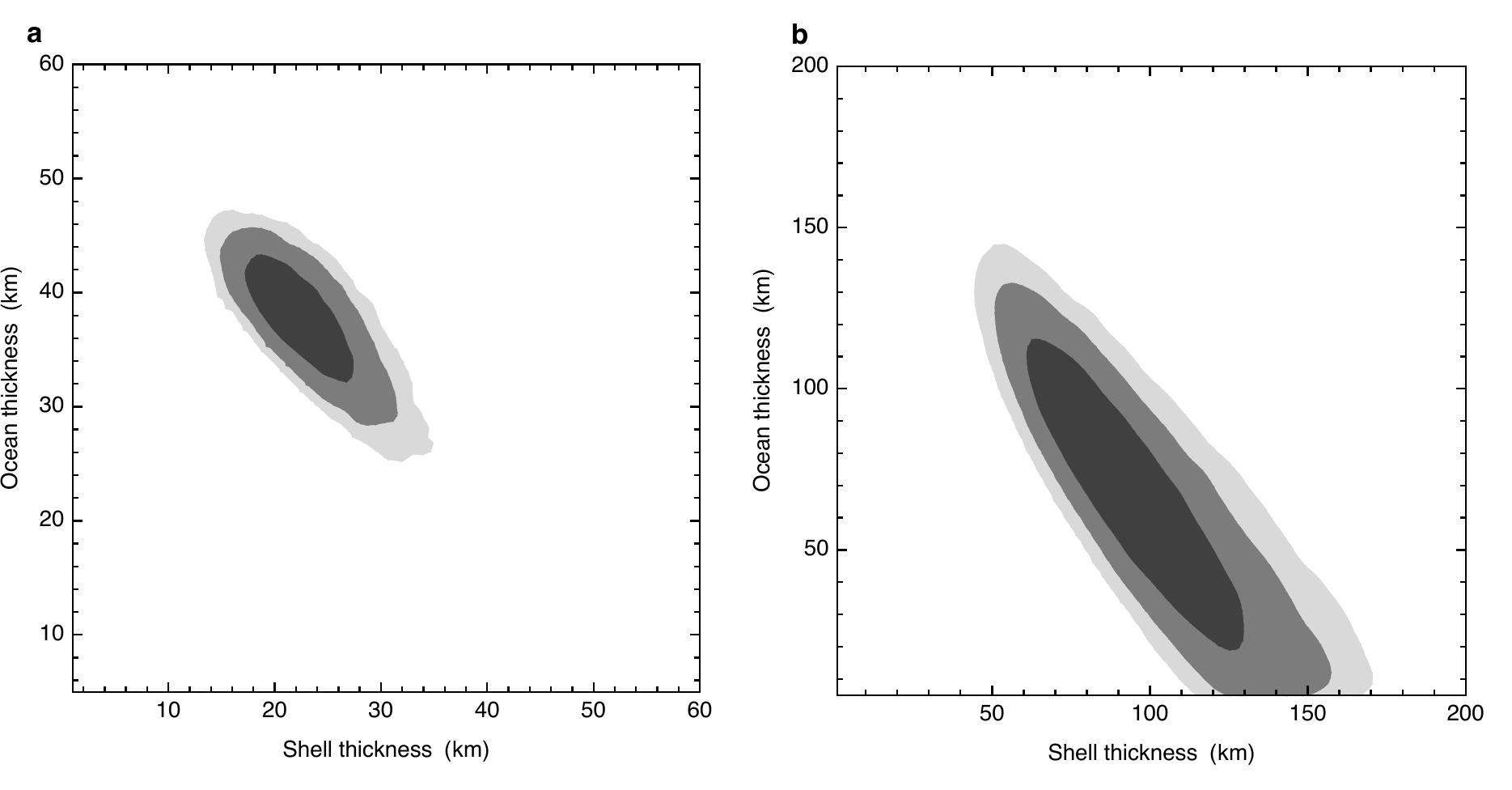}
   \caption[]{
   Inferred shell and ocean thicknesses: (a) Enceladus, (b) Dione.
   Contours show Bayesian confidence regions to $(1\sigma,2\sigma,3\sigma)$.
   The ranges on the axes correspond to the prior uniform distributions.
   The inverse correlation between shell and ocean thicknesses is clearly visible.}
   \label{Fig2}
\end{figure}

\begin{figure}
   \includegraphics[width=14cm]{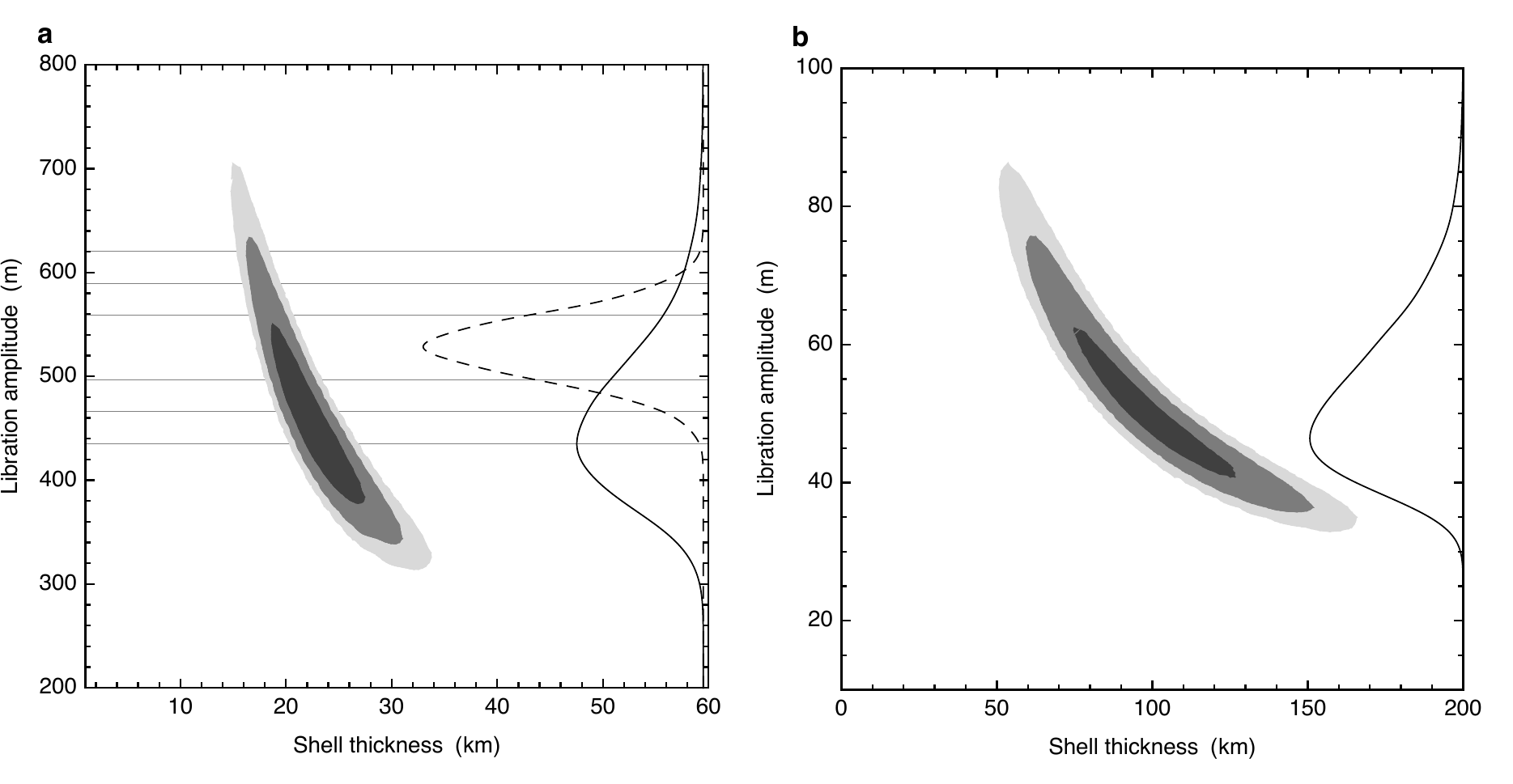}
   \caption[]{
   Libration amplitude: (a) Enceladus, (b) Dione.
   Contours show Bayesian confidence regions for models resulting from the gravity-shape inversion $(1\sigma,2\sigma,3\sigma)$.
   Solid curves show the distributions of inferred librations.
   In panel (a), the dashed curve shows the distribution of observed librations \citep{thomas2016} with horizontal lines indicating the $(1\sigma,2\sigma,3\sigma)$ ranges.}
   \label{Fig3}
\end{figure}

For Enceladus, our reference data are the coefficients $(C_{20},C_{22},C_{30})$ of the gravity solution SOL1 (Table \ref{TableGraviEnc}), combined with the shape TOPA (Table \ref{TableTopoEnc}).
For this model, the inversion yields $d_s=23\pm4\rm\,km$, $d_o=38\pm4\rm\,km$, $r_c=192\pm2\rm\,km$, and $\rho_c=2422\pm46\rm\,kg/m^3$ at $1\sigma$ (see Fig.~\ref{FigHistoSOL1} for posterior distributions).
Shell and ocean densities are not constrained.
The shell and ocean thicknesses are inversely correlated (Fig.~\ref{Fig2}) because the core radius is well determined (Fig.~\ref{FigCorrelCore}).
Our estimates for Enceladus's shell thickness overlap with those of librations ($21-26\rm\,km$ in \citet{thomas2016}, $14-26\rm\,km$ in \citet{vanhoolst2016}).
For a rigorous comparison, we compute the librations from the probability distribution over the parameters inferred from our gravity-shape inversion assuming a rigid shell with nonhydrostatic boundaries \citep{vanhoolst2016}.
The predicted distribution ($461\pm72\rm\,m$ at $1\sigma$) is wider than the distribution of observed librations ($528\pm31\rm\,m$ at $1\sigma$) (Fig.~\ref{Fig3}).
Thus librations put tighter bounds on the average shell thickness of Enceladus, although they do not constrain the other interior parameters.

For Dione, the inversion of gravity and shape data (Tables \ref{TableGraviDione} and \ref{TableTopoDione})  yields $d_s=99\pm23\rm\,km$, $d_o=65\pm30\rm\,km$, $r_c=398\pm14\rm\,km$, and $\rho_c=2435\pm140\rm\,kg/m^3$ (at $1\sigma$).
Errors are much larger than for Enceladus (Fig.~\ref{Fig2}) because of the large relative error on the shape (Tables \ref{TableTopoDione} and \ref{TableInvDione}).
The gravity and shape data can thus be explained if there is a global ocean deep under the surface, whose past existence was already suggested by observation of ridge flexure \citep{hammond2013} and thermal history models \citep{multhaup2007}.
We predict that Dione undergoes librations of amplitude $52\pm10\rm\,m$ at $1\sigma$ (Fig.~\ref{Fig3}), one order of magnitude below Enceladus and thus not detectable in Cassini images.

\section*{5 Discussion}

Comparison with librations can pinpoint data biases and constrain modeling choices.
First, Enceladus's degree-three gravity favors a thinner shell than degree-two coefficients \citep{mckinnon2015} (Fig.~\ref{Fig1} and Table \ref{TableInvEnc}).
Degree-two gravity however predicts a thinner shell (in agreement with librations) if $C_{22}$ is $2\sigma$ higher than its SOL1 central value, as suggested by the alternative gravity solution SOL2 (Table \ref{TableGraviEnc}).
Alternatively, the degree-two shape could be responsible for the disagreement: while the new ellipsoidal shape of \citet{thomas2016} does not affect much the results (Table \ref{TableInvEnc}), the recent ellipsoidal shape of \citet{nadezhdina2016} predicts more degree-two compensation and thus a thinner shell.
Second, we did not allow for a lot of surface porosity in our three-layer model, but we can easily do it with a four-layer model.
The estimated shell thickness increases with porosity because surface topography contributes less to the gravity signal and must be less compensated (Table \ref{TableInvEnc} and Fig.~\ref{FigPorous}).
Consistency with librations suggests however that porosity is not an important factor.

\begin{figure}
   \centering
   \includegraphics[width=7cm]{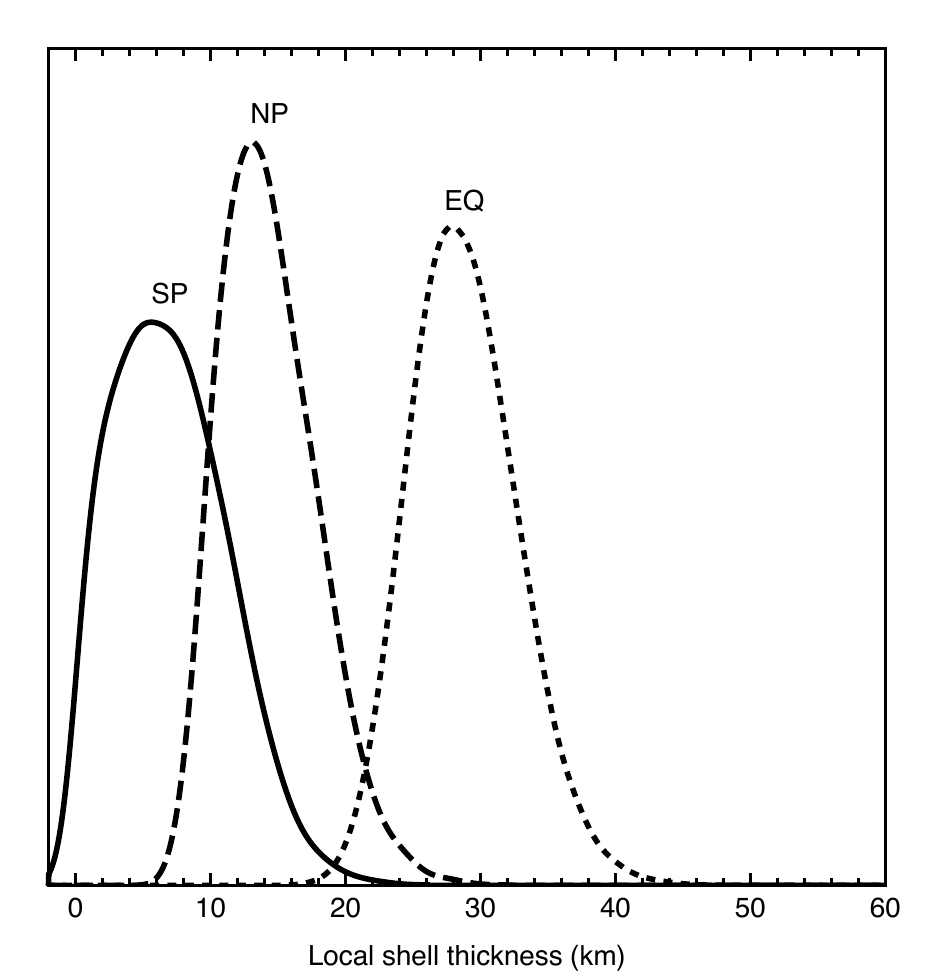}
   \caption[]{
   Local shell thickness of Enceladus.
   Solid, dashed, and dotted curves show the distributions of the inferred shell thickness at the south pole (SP), north pole (NP) and equator (EQ), respectively.
   The equatorial thickness represents the zonal average.
   Negative values are an artefact of histogram smoothing.
   }
   \label{Fig4}
\end{figure}

Enceladus's shell thickness varies mainly in latitude from $29\pm4\rm\,km$ at the equator (zonal average) to $14\pm4\rm\,km$ and $7\pm4\rm\,km$ at the north and south poles, respectively (Fig.~\ref{Fig4} and Text \ref{TextS5}).
Longitudinal variations are either subdominant, with shell thickening along the tidal axis, or could be absent altogether as suggested by the latest estimates of the ellipsoidal shape (Table \ref{TableTopoEnc}).
The very thin south polar crust facilitates the passage of water from the ocean to the surface and increases the concentration of tidal heating in the area.
The variation in shell thickness could be due to crustal tidal heating which is indeed highest at the poles and lowest along the tidal axis \citep{ojakangas1989a}.
For Dione, shell thickness varies by less than five percent with a minimum at the poles and a maximum along the leading-trailing axis.
Similarly to Enceladus, the zonal variation in shell thickness could be due to tidal heating, but we have no ready explanation for the longitudinal variation anticorrelated with tidal heating.
Both satellites are more tectonized or resurfaced in their leading and trailing hemispheres than close to their tidal axis \citep{crowwillard2015,kirchoff2015}.

The average crustal stress is about $30\rm\,kPa$ for Enceladus, which is half of the average topographic stress, as expected in isostasy \citep{melosh2011}.
It is comparable in magnitude to tidal stresses \citep{nimmo2007b} and could trigger the formation of the south polar terrain margins by gravitational spreading \citep{yin2015}.
We also evaluated isostatic stresses in the core in order to check that they are always smaller than isostatic stresses in the crust.
Therefore, the core can be modeled as an elastic layer regarding isostatic loading.
A final contentious point is that isostasy is only valid to first order in the flattening, contrary to the figure of equilibrium.
The second-order error on the isostatic model, however, changes the total gravity potential \citep{wieczorek1998} by less than the data uncertainty (Text \ref{TextS6}).

Beyond Enceladus and Dione, our new take on isostasy is applicable to large icy satellites with global oceans, such as Europa \citep{nimmo2007}, Titan \citep{nimmo2010}, and particularly Ganymede whose gravity and shape will be measured by the JUICE mission.
According to \citet{park2016}, gravity and shape data from the Dawn mission suggest isostasy on Ceres, but the case is far from clear because compensation does not occur for all gravity components.
Finally, isostasy plays a crucial role in understanding the long-wavelength gravity and shape as well as estimating the crust thickness of the planets Mars \citep{wieczorek2004}, Venus \citep{james2013}, and Mercury \citep{perry2015}.
Thanks to the simultaneous availability of gravity/shape and libration data, Enceladus's case constitutes the first validation of planetary-scale isostasy.

\section*{Acknowledgments}
\small
All data used here are freely available in the literature.
M.B. is supported by the Brain Pioneer contract BR/314/PI/LOTIDE.
A.R. is supported by the Belgian PRODEX program managed by the European Space Agency in collaboration with the Belgian Federal Science Policy Office.
A.T. received support from the `Supplementary Researcher' programme managed by the Belgian Federal Science Policy Office, and from the European Research Council (ERC) under the European Union's Horizon 2020 research and innovation programme (grant agreement No 670874).

\newpage

\bibliographystyle{agufull04}
\newpage


\renewcommand{\thesection}{S.\arabic{section}}
\setcounter{section}{0}
\setcounter{page}{1}
\setcounter{figure}{0}
\renewcommand{\thetable}{S.\arabic{table}}
\renewcommand{\thefigure}{S.\arabic{figure}}

\par\noindent
{\LARGE SUPPLEMENTARY INFORMATION}\\

\par\noindent
{\LARGE Enceladus's and Dione's floating ice shells \vspace{2mm}\\
supported by minimum stress isostasy}

\tableofcontents
\listoffigures
\listoftables

\newpage

\renewcommand{\theequation}{S.\arabic{equation}} 
\setcounter{equation}{0}  

\section{Nonhydrostatic deviations  of Enceladus and Dione}
\label{TextS1}

The degree-two shape and gravity ratios are defined by \citep{zharkov1985,nimmo2011}
\begin{equation}
F_2 = - \frac{H_{20}}{H_{22}}
\hspace{5mm} \mbox{and} \hspace{5mm}
G_2 = - \frac{C_{20}}{C_{22}} = \frac{J_2}{C_{22}} \, ,
\label{F2}
\end{equation}
where $H_{nm}$ and $C_{nm}$ denote the cosine real harmonic coefficients of degree $n$ and order $m$ of the shape and nondimensional gravity potential, respectively.
If the body is in hydrostatic equilibrium, $F_2=G_2=10/3$ to first order in the flattening, whatever the density stratification.
Second-order corrections change the shape and gravity ratios of a hydrostatic body by up to a few percent with respect to 10/3, with a slight dependence on the internal structure.
The degree-two shape and gravity ratio resulting from Eqs.~(\ref{HstratM})-(\ref{CstratM}) are
\begin{equation}
F_2 = \frac{10}{3} \left( 1 - \frac{48}{35}  \, h_2^F \, q \right)
\hspace{5mm} \mbox{and} \hspace{5mm}
G_2 = \frac{10}{3} \left( 1 - \frac{16}{7}  \, h_2^F \, q \right) .
\end{equation}
If the body is homogeneous, $(F_2,G_2)=(3.26,3.21)$ for Enceladus and $(F_2,G_2)=(3.31,3.30)$ for Dione.
Cassini observations yield $(F_2,G_2)=(4.2,3.5)$ for Enceladus (Tables~\ref{TableGraviEnc} and \ref{TableTopoEnc}) and $(F_2,G_2)=(5.2,4.0)$ for Dione (Tables~\ref{TableGraviDione} and \ref{TableTopoDione}).
To second order in the flattening, the nonhydrostatic shape deviations
(with respect to the homogeneous body) are thus $\delta{}F_2=29\%$ for Enceladus and $\delta{}F_2=58\%$ for Dione, while the nonhydrostatic gravity deviations
are $\delta{}G_2=9\%$ for Enceladus and $\delta{}G_2=21\%$ for Dione.
The density stratification decreases these values by less than one percent ($\delta{}F_2=28\%$ and $\delta{}G_2=8\%$ for Enceladus if $k_2^F=0.93$).

\section{Gravity data (Tables~\ref{TableGraviEnc} \& \ref{TableGraviDione})}
\label{TextS2}

Two gravity solutions are available for Enceladus \citep{iess2014}.
SOL1 includes the five coefficients of degree two and the zonal coefficient of degree three, i.e.\ the least set of parameters able to fit the data at the noise level and compatible with the topography.
SOL2 includes the five coefficients of degree two and the seven coefficients of degree three.
Given the uncertainties, SOL2 does not contain more physical information than SOL1.
On the other hand, the errors on SOL1 cannot reflect all the biases due to the sparse data (only three flybys) or resulting from the assumptions involved in spacecraft trajectory modeling (discussed in the Supplementary Material of \citet{iess2014}).
In particular, the central value of $C_{22}$ changes from SOL1 to SOL2 by more than $4\sigma$ while $C_{30}$ changes by more than $1\sigma$ and becomes compatible with a zero value within $2\sigma$ (the shift in $C_{20}$ is negligible).
The error bars being much larger on SOL2, the shift of $C_{22}$ may be meaningless.
Nevertheless we retain both gravity solutions in order to assess the impact of gravity field uncertainties on interior modeling.
Table~\ref{TableGraviEnc} gives the coefficients differing from zero at $3\sigma$ (at least for SOL1): $C_{20}$, $C_{22}$, and $C_{30}$.
Table~\ref{TableGraviDione} gives the coefficients of the preliminary gravity solution for Dione \citep{hemingway2016}.

The geoid, to first order in the figure of equilibrium, is given by $H_{nm}^h = R_0 C_{nm}$, except for the two components affected by the rotation and by the permanent static tide \citep{zharkov2004,zharkov2010}:
\begin{equation}
H_{20}^h/R_0 = C_{20} - \frac{5}{6} \, q
\hspace{5mm} \mbox{and} \hspace{5mm}
H_{22}^h/R_0 = C_{22} + \frac{1}{4} \, q \, .
\label{geoid}
\end{equation}
The second-order geoid is given by Eq.~(50) of \citet{zharkov2010} but second-order corrections are only a few meters.

\section{Shape data (Tables~\ref{TableTopoEnc} \& \ref{TableTopoDione})}
\label{TextS3}

The shape of the mid-size Saturnian satellites has been determined by fitting an ellipsoid  \citep{thomas2007,thomas2016,nadezhdina2016} and by directly estimating the spherical harmonic components of the topography \citep{nimmo2011}.
For Enceladus and Dione, we use the shape of \citet{nimmo2011} (denoted TOPA), which provides all harmonic coefficients of degree two and three.
For Enceladus, we also use the ellipsoidal shape of \citet{thomas2016} (denoted TOPB), which is based on more recent data, from which the harmonic coefficients of degree two can be derived. 
TOPB is less flattened than TOPA and its coefficient $H_{22}$ follows the geoid (see Table~\ref{TableTopoEnc}).
The new ellipsoidal shape of \textit{Nadezhdina et al.} [2016] has a polar flattening ($-3H_{20}/2R$) close to the one of TOPA but its equatorial flattening ($6H_{22}/R$) is nearly hydrostatic (similarly to TOPB).

We use here spherical harmonic coefficients in unnormalized form (as in \citet{iess2014,mckinnon2015}) so that $H_{nm}$ denote the coefficients of $P_{nm}(\cos\theta)\cos{}m\phi$ (error bars on $H_{nm}$ are not correct in \citet{iess2014} but are correctly given in \citet{mckinnon2015}).
\citet{thomas2016} give estimates for the semi-major axes $a>b>c$ and the mean radius $R_0$ of the best-fitting ellipsoid with $2\sigma$ error bars;
the mean radius is defined as the radius of the sphere of equal volume; its value in \citet{thomas2016} should be $252.0\rm\,km$ (P.~Thomas, private communication).
The degree-two harmonic coefficients of the shape, to first order in the flattening, are computed from
\begin{equation}
H_{20} = -\frac{2}{3} \left( \frac{a+b}{2} -c\right)
\hspace{5mm} \mbox{and} \hspace{5mm}
H_{22} = \frac{a-b}{6} \, .
\end{equation}

\section{Basic equations of Bayesian inversion}
\label{TextS4}

The result of the Bayesian inversion is the posterior probability density function, i.e.\ the conditional probability for the parameters $\mathcal{P}$ given the data $\mathcal{D}$ \citep{tarantola2005}:
\begin{equation}
	\mathrm{p} (\mathcal{P} | \mathcal{D}) =
	\frac{ \mathcal{L}(\mathcal{D}|\mathcal{P}) \, \Pi(\mathcal{P}) }{\int d\mathcal{P} \; \mathcal{L}(\mathcal{D}|\mathcal{P})\, \Pi(\mathcal{P}) } \, .
\end{equation}
$\Pi$ represents the prior information on $\mathcal{P}$ while $\mathcal{L}$ is the likelihood function associating a probability to the data given parameter values.
$\mathcal{P}$ is related to $\mathcal{D}$ by Eqs.~(\ref{decompH}) to (\ref{CstratM}) and (\ref{fn}), though we replace in practice Eqs.~(\ref{HstratM})-(\ref{CstratM}) by the complete second-order solutions for the figure of equilibrium.
We assume that the gravity data ($C_{20},C_{22},C_{30}$) are normally distributed and uncorrelated (Tables~\ref{TableGraviEnc} and \ref{TableGraviDione}).
The model parameters are the densities ($\rho_s$, $\rho_o$) and thicknesses ($d_s$, $d_o$) of the ice shell and ocean (the radius $r_c$ and density $\rho_c$ of the core are derived parameters).
The prior information on $\mathcal{P}$ is described by uniform probability density functions (ranges given in Section~3.4) subjected to the constraint that the shell thickness at the south pole is not negative (Fig.~\ref{FigHistoSOL1} shows the resulting prior distributions).

If the data $\mathcal{D}$ includes the shape coefficients ($H_{20},H_{22},H_{30}$), the isostatic shape coefficients must be treated as additional model parameters.
We choose instead to include the shape coefficients with their uncertainties in the model itself which becomes probabilistic.
This is similar to the problem of fitting a straight line to $(x,y)$ data with errors in both coordinates.
According to Section~4.8.2 of \citet{gregory2005}, the likelihood function is the same as if the shape was exactly known except that the standard deviation of the gravity data is replaced by
\begin{equation}
\sigma_{nm}^2 = \sigma_{Cnm}^2 + \left( Z_n \, \sigma_{Hnm} \right)^2 \, ,
\end{equation}
where the $\sigma_{Cnm}$ and $\sigma_{Hnm}$ are the $1\sigma$ uncertainties on the gravity (Tables~\ref{TableGraviEnc} and \ref{TableGraviDione}) and shape (Tables~\ref{TableTopoEnc} and \ref{TableTopoDione}) coefficients, respectively.
The likelihood function $\mathcal{L}$ reads
\begin{equation}
\mathcal{L}(\mathcal{D}|\mathcal{P}) = \mathcal{L}_{20} \, \mathcal{L}_{22} \, \mathcal{L}_{30} \, ,
\end{equation}
where
\begin{equation}
\mathcal{L}_{nm} = \frac{1}{\sqrt{2\pi}\sigma_{nm}} \exp -\frac{\left( C_{nm}^h + Z_n ( H_{nm} - H_{nm}^h ) - C_{nm} \right)^2}{2\sigma_{nm}^2} \, ,
\end{equation}
it being understood that $H_{30}^h=C_{30}^h=0$.

We simulate $\mathrm{p}$ with a Metropolis-Hastings sampler \citep{tarantola2005}, generating 400000 samples for each inversion and retaining every fourth sample in order to reduce the correlation within the samples.
The burn-in length is equal to 1000.
From the selected samples, we compute for each parameter the probability density function (Fig.~\ref{FigHistoSOL1}), the mean value and the Bayesian confidence intervals (Tables~\ref{TableInvEnc} and \ref{TableInvDione}).
The correlation between parameters is illustrated by Fig.~2 showing the two-parameter Bayesian confidence regions for the ice shell and ocean thicknesses, while Fig.~\ref{FigCorrelCore} does the same for the core radius and density.\\

\section{Bayesian inversion results (Tables~\ref{TableInvEnc} \& \ref{TableInvDione})}
\label{TextS5}

\begin{figure}
   \centering
   \includegraphics[width=4.8cm]{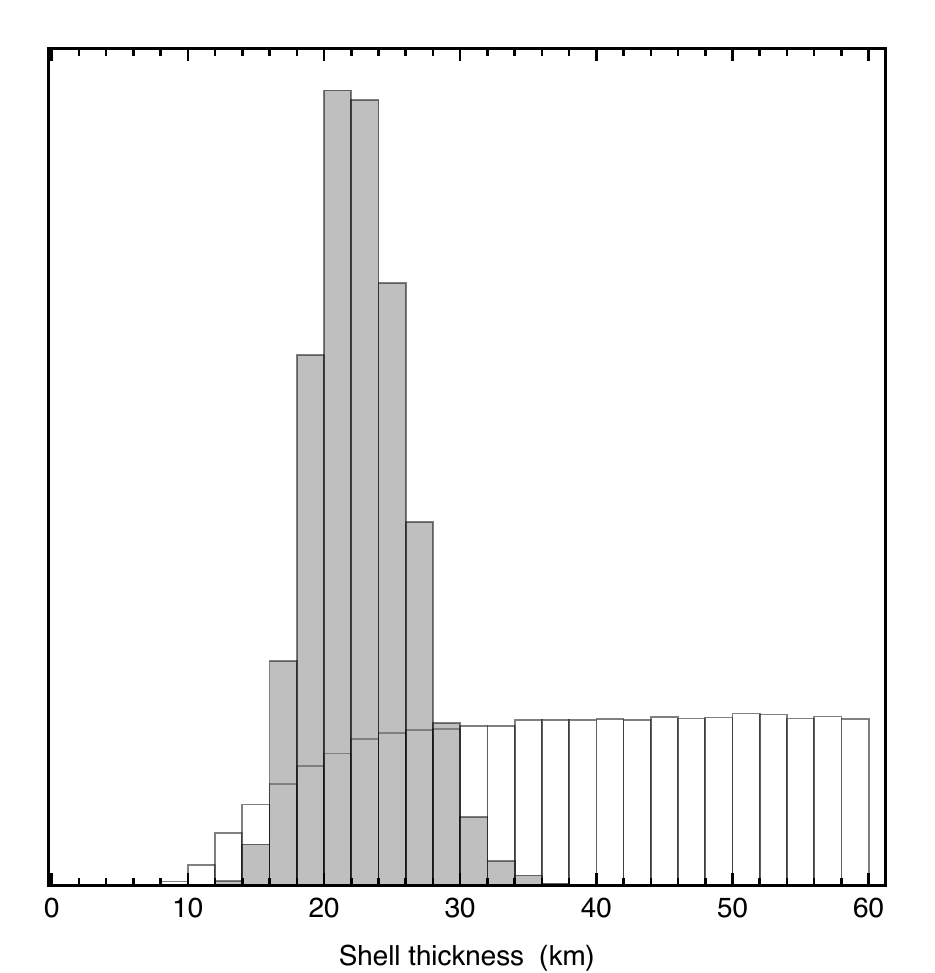}
   \includegraphics[width=4.8cm]{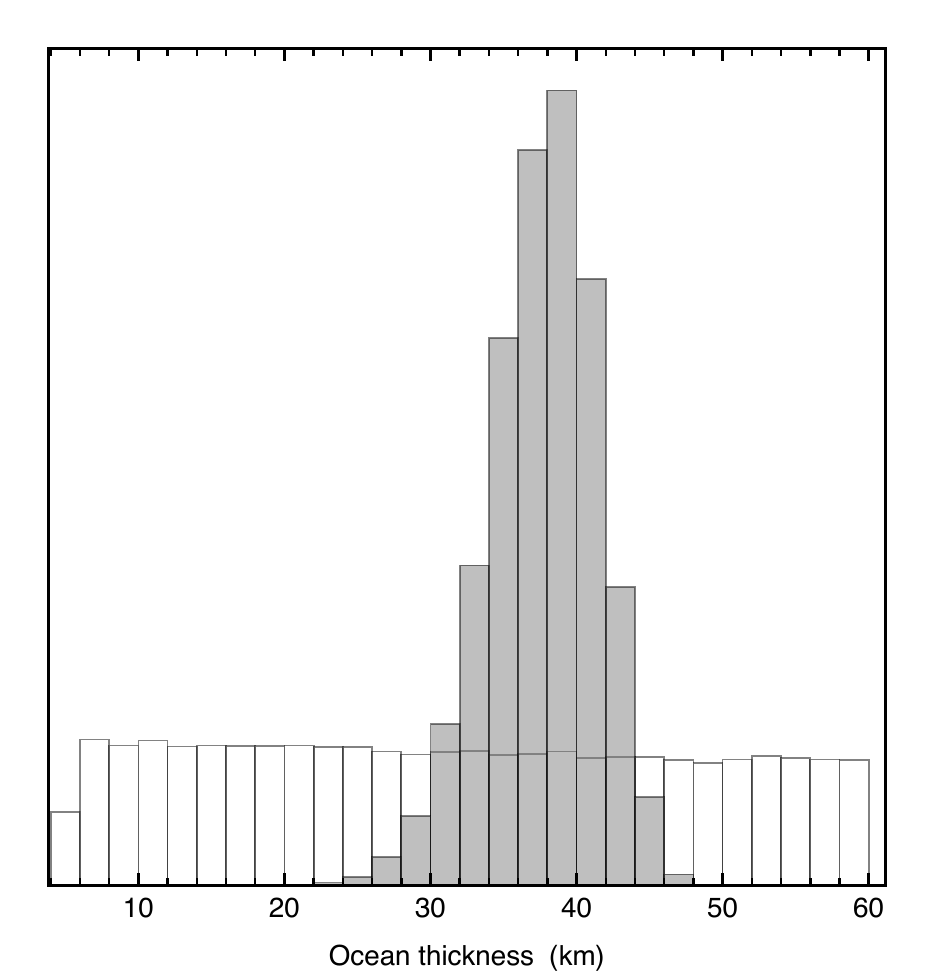}
   \includegraphics[width=4.8cm]{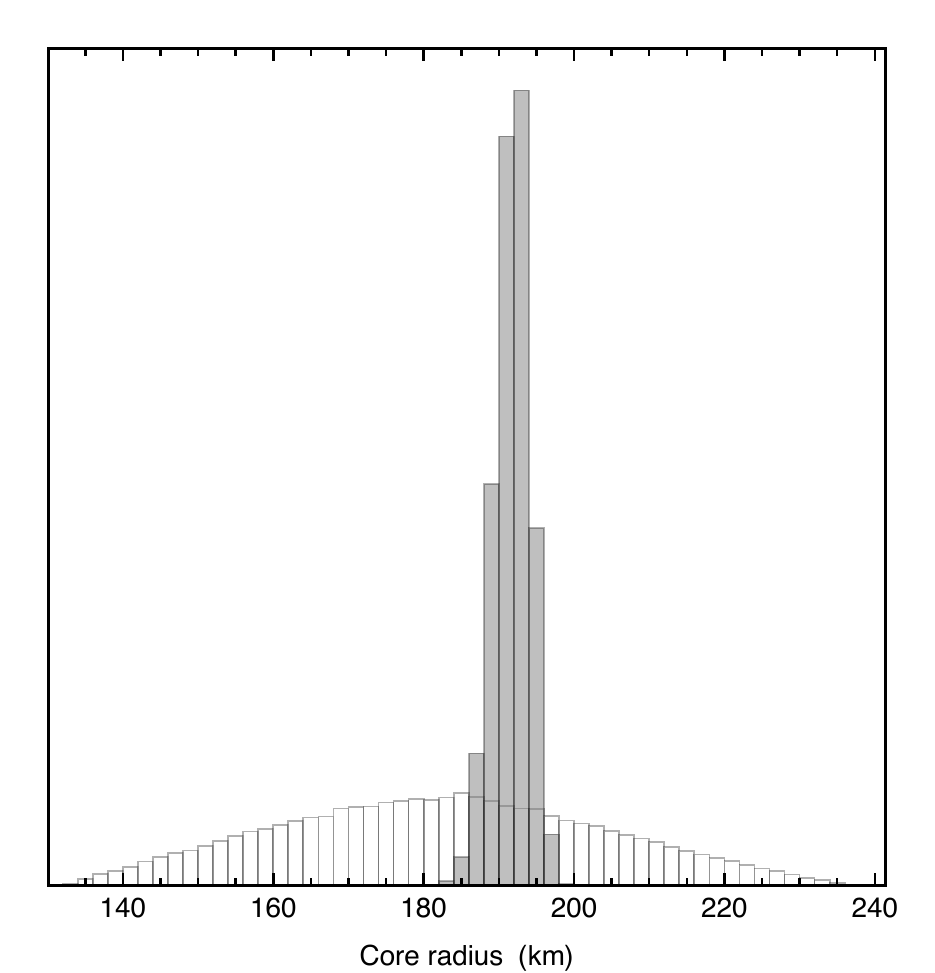}
   \includegraphics[width=4.8cm]{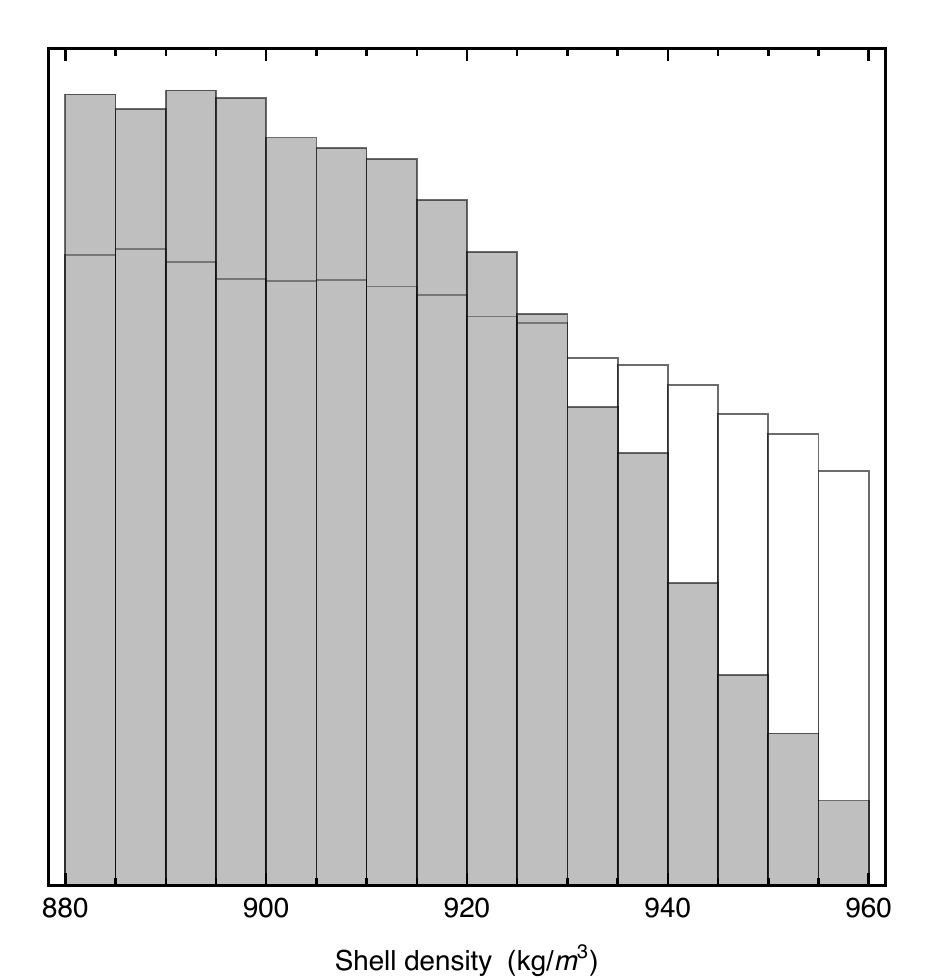}
   \includegraphics[width=4.8cm]{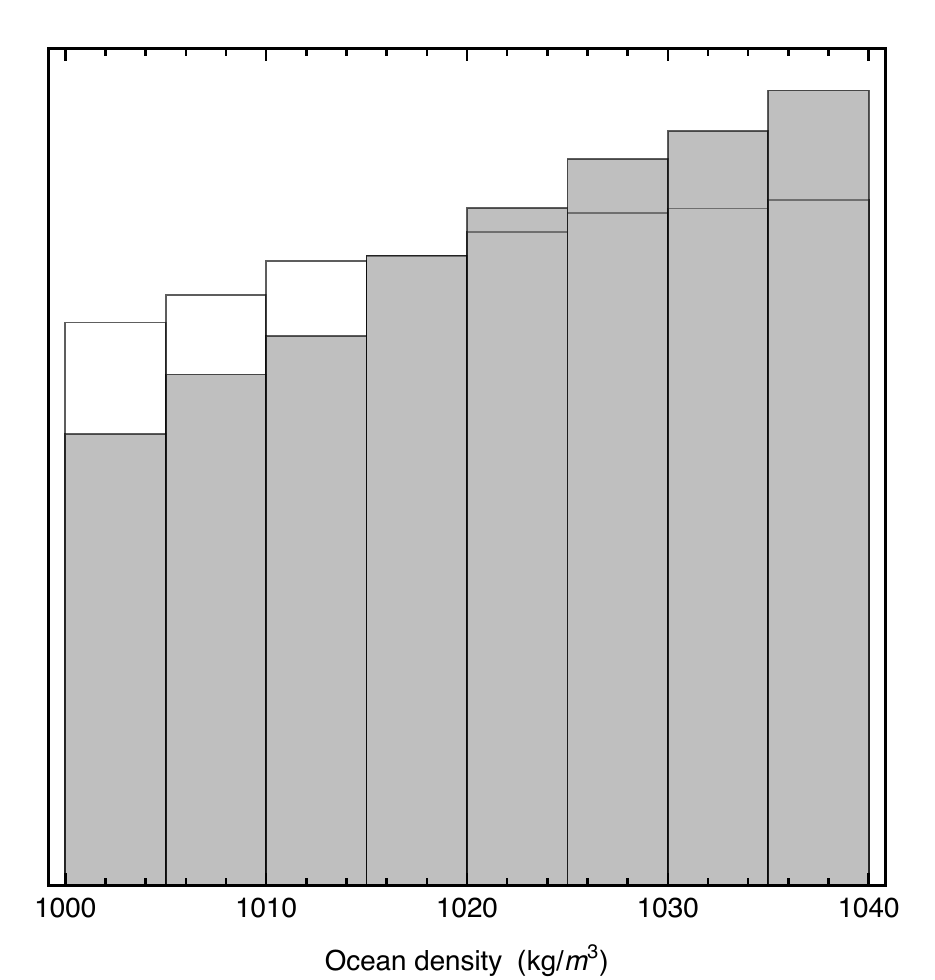}
   \includegraphics[width=4.8cm]{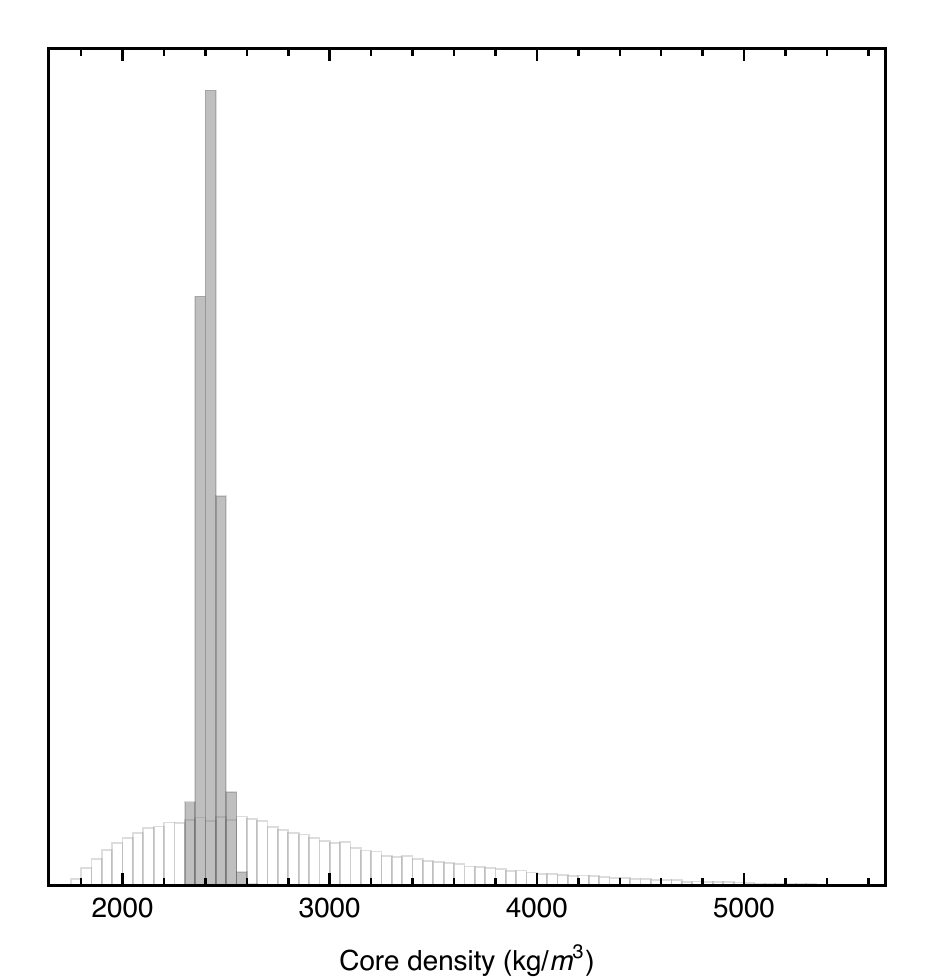}
   \includegraphics[width=4.8cm]{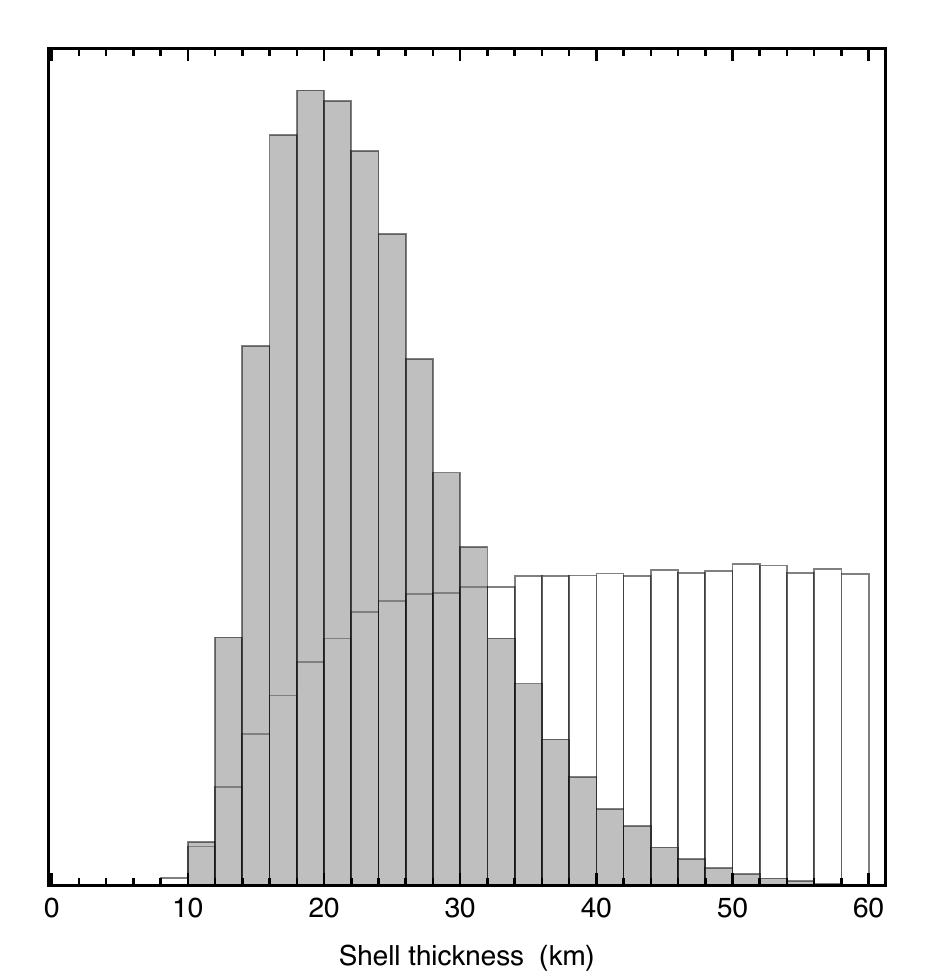}
   \includegraphics[width=4.8cm]{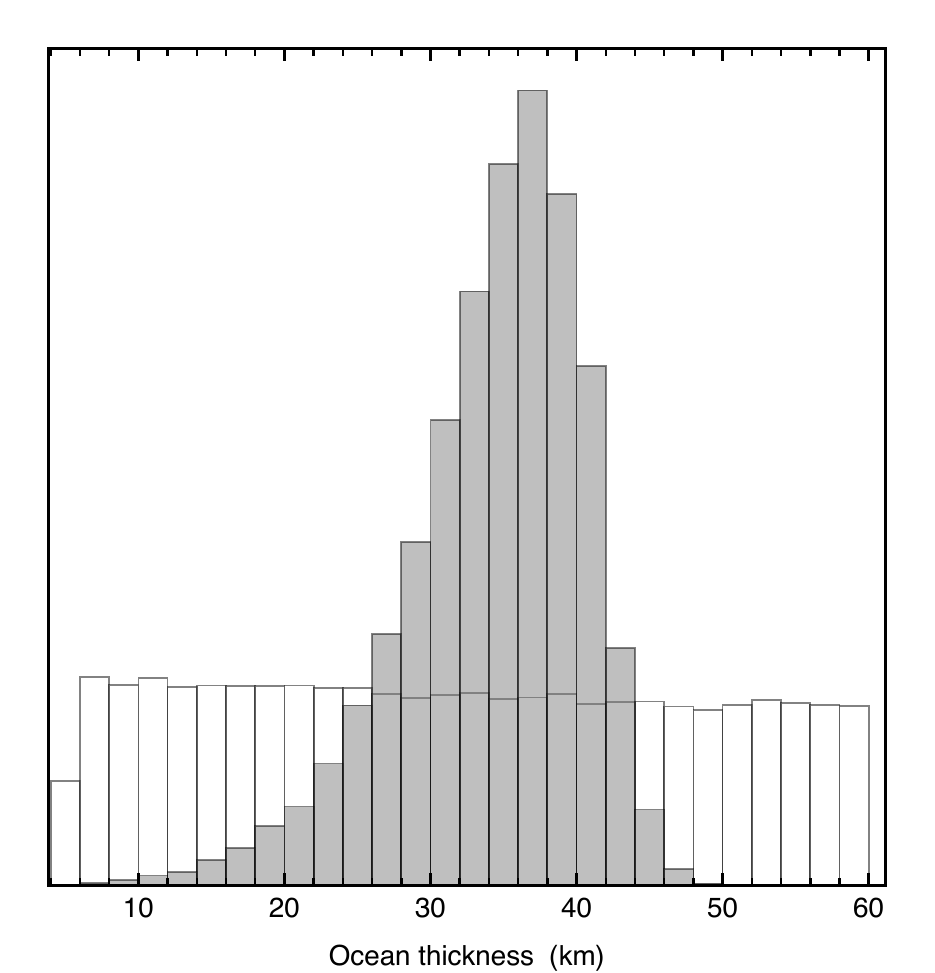}
   \includegraphics[width=4.8cm]{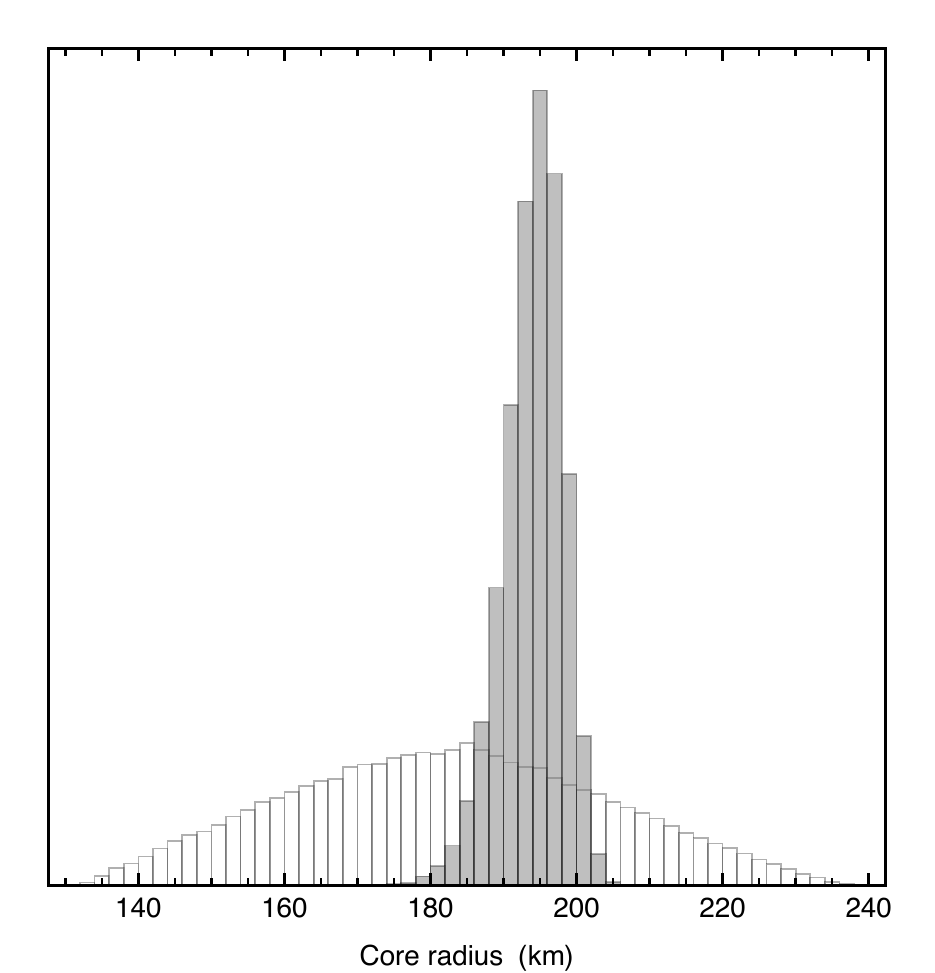}
   \includegraphics[width=4.8cm]{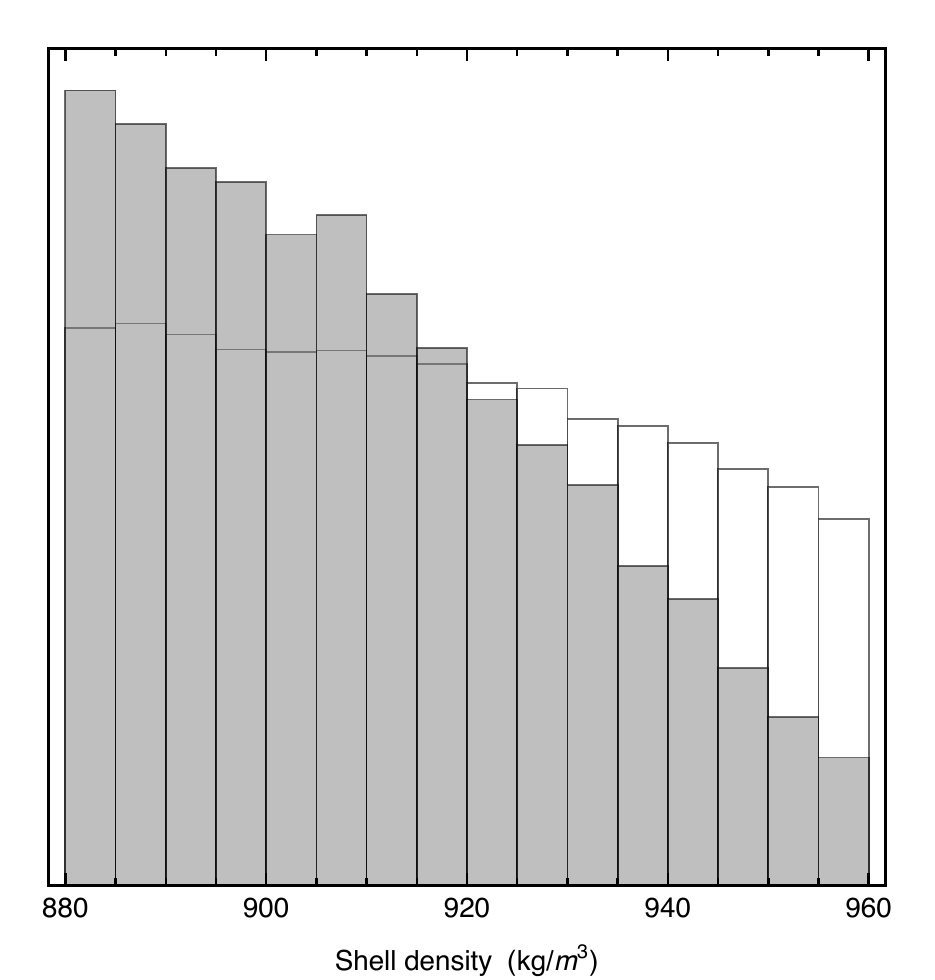}
   \includegraphics[width=4.8cm]{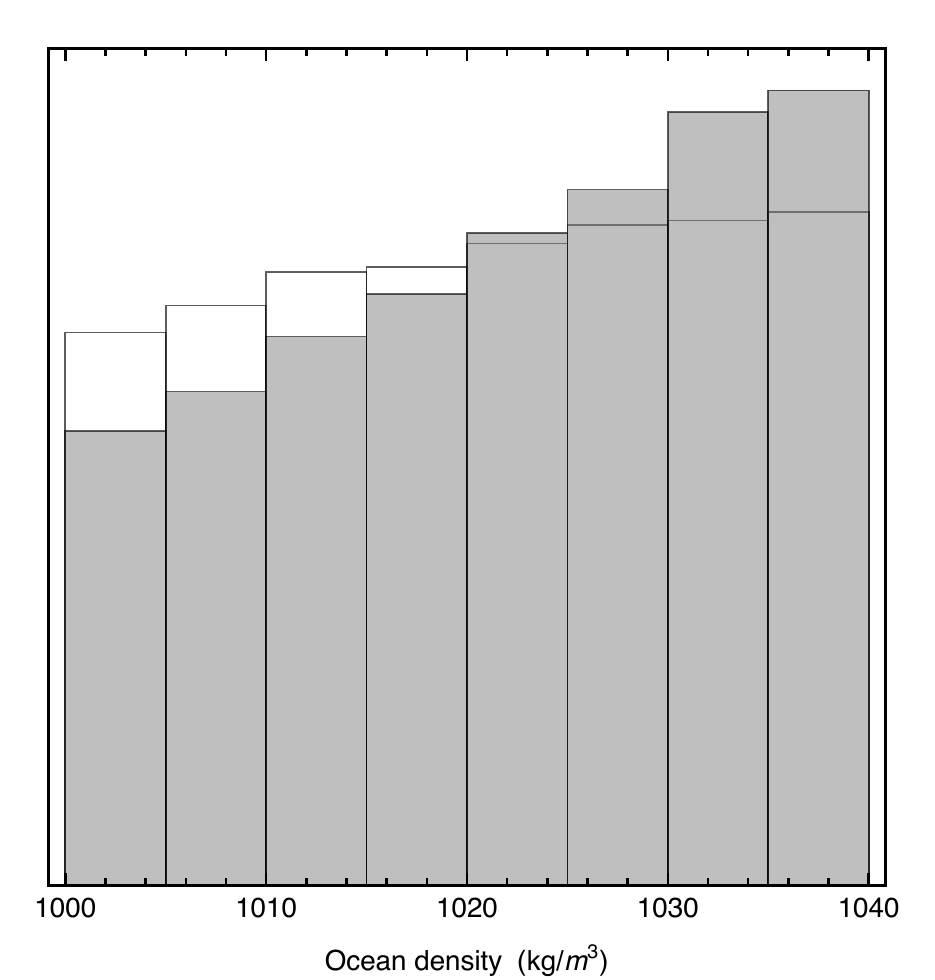}
   \includegraphics[width=4.8cm]{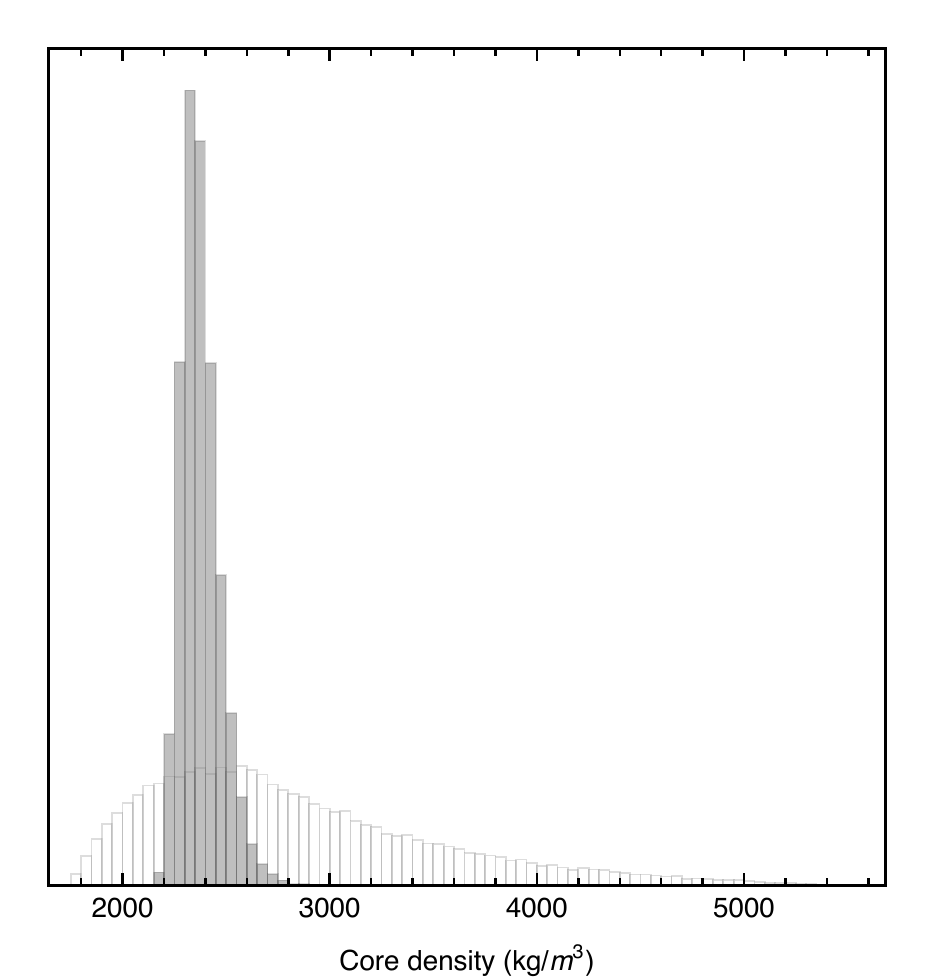}
   \caption[Prior and posterior distributions of interior parameters for Enceladus]{
   Prior and posterior distributions of interior parameters for Enceladus: thickness and density of the crust (or ice shell), ocean, and core.
   The models are SOL1/TOPA (top six panels) and SOL2/TOPA (bottom six panels).
   Priors are transparent and posteriors are shaded.
   Priors are the same in the two models.
   }
   \label{FigHistoSOL1}
\end{figure}

\begin{figure}
  \includegraphics[width=14cm]{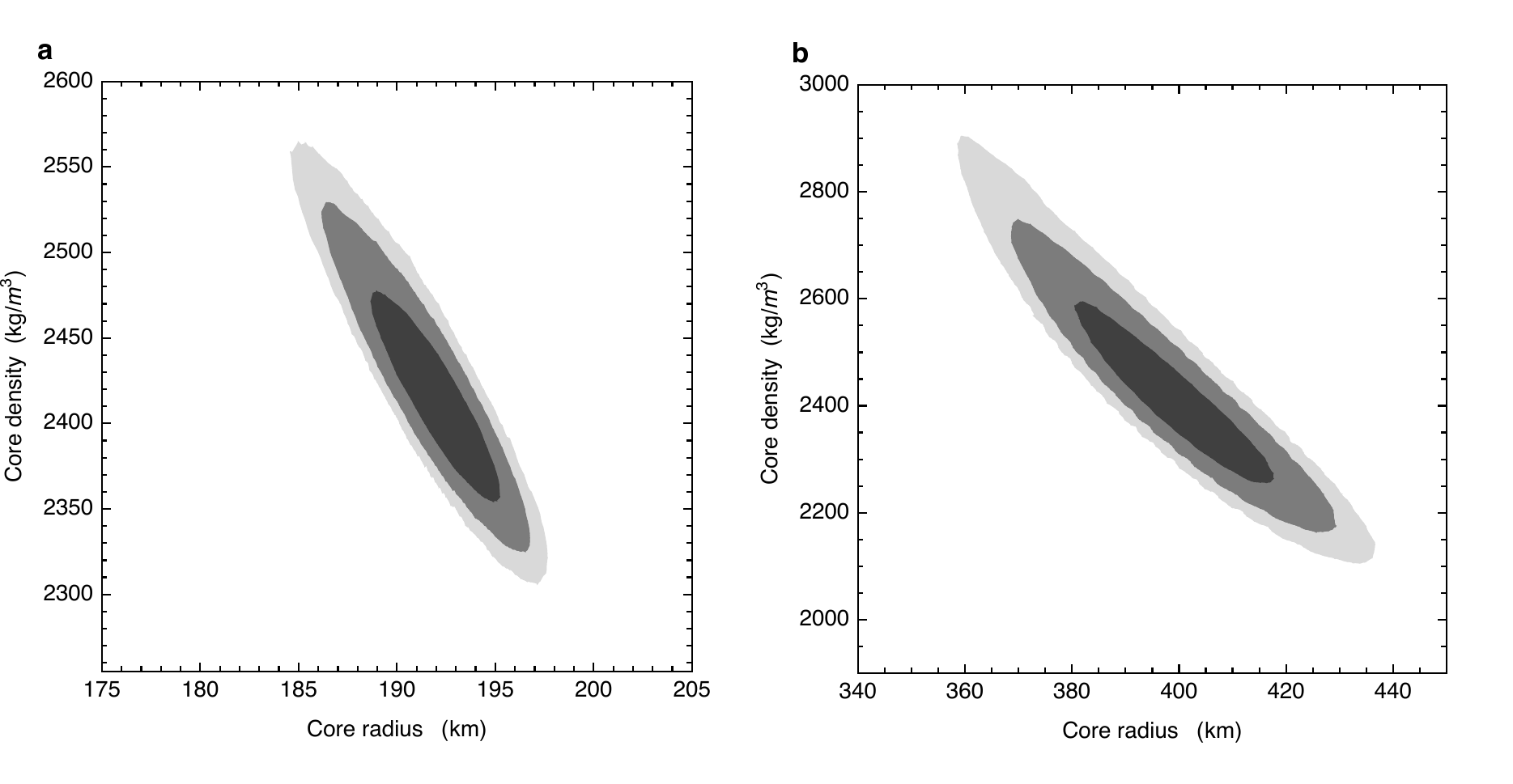}
   \caption[Bayesian confidence regions for core radius and core density]{
   Bayesian confidence regions for core radius and core density. (a) Enceladus, (b) Dione.
   Contours show Bayesian confidence regions to $(1\sigma,2\sigma,3\sigma)$ for the model SOL1/TOPA.
   The inverse correlation between the core radius and core density is clearly visible.}
   \label{FigCorrelCore}
\end{figure}

Table~\ref{TableInvEnc} shows the results of several inversions of gravity-shape data for Enceladus, corresponding either to different datasets or to different interior models.
SOL1 and SOL2 refer to the gravity data of Table~\ref{TableGraviEnc} while TOPA and TOPB refer to the shape data of Table~\ref{TableTopoEnc}.
The SOL1/TOPA column shows the results of the reference case discussed in the text (Fig.~\ref{FigHistoSOL1} shows the prior and posterior distributions).
The SOL2/TOPA column shows the results with the alternative gravity solution SOL2.
In that case, the mean value of the shell thickness ($d_s=24\rm\,km$) is slightly higher than with SOL1 ($d_s=23\rm\,km$) because the larger data uncertainties lead to a distribution extending to much larger values of the shell thickness, though the mode is $d_s=20\rm\,km$ (Fig.~\ref{FigHistoSOL1}).

\begin{figure}
   \centering
   \includegraphics[width=7cm]{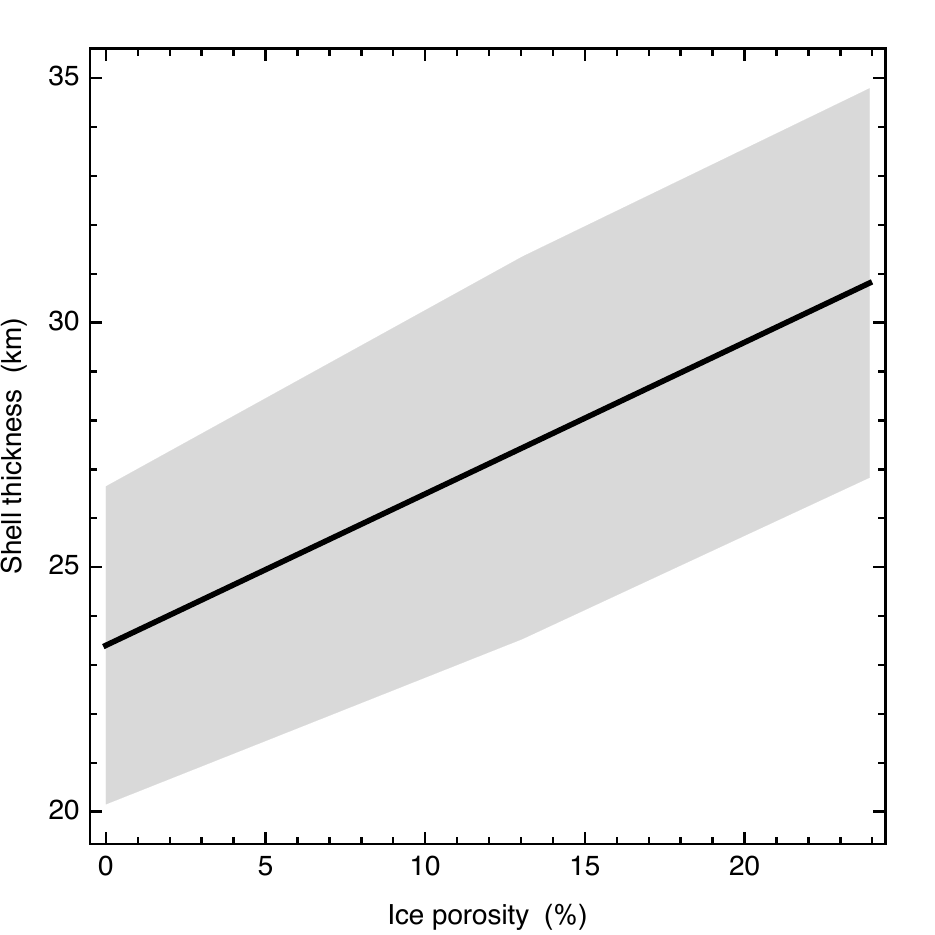}
   \caption[Enceladus's shell thickness if porous crust]{
   Enceladus's shell thickness if porous crust.
   The crust is split in a bottom layer with a density of $920\rm\,kg/m^3$ and a $10\rm\,km$-thick upper layer with a porosity varying between 0\% ($920\rm\,kg/m^3$) and 24\% ($700\rm\,kg/m^3$).
   The solid curve shows the mean value resulting from the Bayesian inversion while the shaded area shows the $1\sigma$ confidence interval.
   Several inversions were made for given values of the porosity, contrary to the POROUS model of Table~\ref{TableInvEnc} in which near-surface porosity is a free parameter.
   }
   \label{FigPorous}
\end{figure}

The next two columns of Table~\ref{TableInvEnc} show the results with different shape data.
The shape given by TOPB is only of degree two, thus we compare inversions for SOL1/TOPA and SOL1/TOPB without the third degree.
The last column of Table~\ref{TableInvEnc} shows the results for an alternative interior model (POROUS) in which the near-surface porosity is a free parameter (see Section~3.4).
As expected, the estimated shell thickness is significantly higher because the surface topography contributes less to the gravity coefficients.
Fig.~\ref{FigPorous} shows the dependence of the estimated shell thickness on crustal porosity (for that figure, inversions were made for given values of the porosity).

Table~\ref{TableInvDione} shows the results of the inversion of gravity-shape data for Dione.
SOL1 and TOPA refer to the gravity and shape data of Tables~\ref{TableGraviDione} and \ref{TableTopoDione}.
The important effect of the shape uncertainties is illustrated by an inversion in which the shape is precisely known (TOPB).

The distribution of the local shell thickness at the colatitude $\theta$ and longitude $\phi$ is obtained by computing for each sample
\begin{equation}
d_{loc}(\theta,\phi) = d_s + \sum_{n=2,3} H_{n0}^{crust} \, P_{n0}(\cos\theta) + H_{22}^{crust} \,  P_{22}(\cos\theta) \, \cos2\phi \, ,
\end{equation}
where $H_{nm}^{crust}=H_{nm}^{iso}-H_{nm}^{int}$ and $P_{nm}$ are the unnormalized associated Legendre functions.
For Enceladus's model SOL1/TOPA, the parameters in the right-hand side have mean values and $1\sigma$ uncertainties given by $d_s=22.8\pm4\rm\,km$, $H_{20}^{crust}=-12.1\pm2.4\rm\,km$, $H_{22}^{crust}=1.3\pm0.3\rm\,km$, and $H_{30}^{crust}=3.7\pm0.7\rm\,km$.
These parameters are correlated so that the error on $d_{loc}(\theta,\phi)$ can only be estimated from the complete distribution.

\begin{figure}

 \hspace{-3cm}
\begin{subfigure}[b]{.5\linewidth}
\captionsetup{skip=-0.cm}
\caption{}\hspace{3.9cm}\label{fig:errora}
\includegraphics[width=6cm]{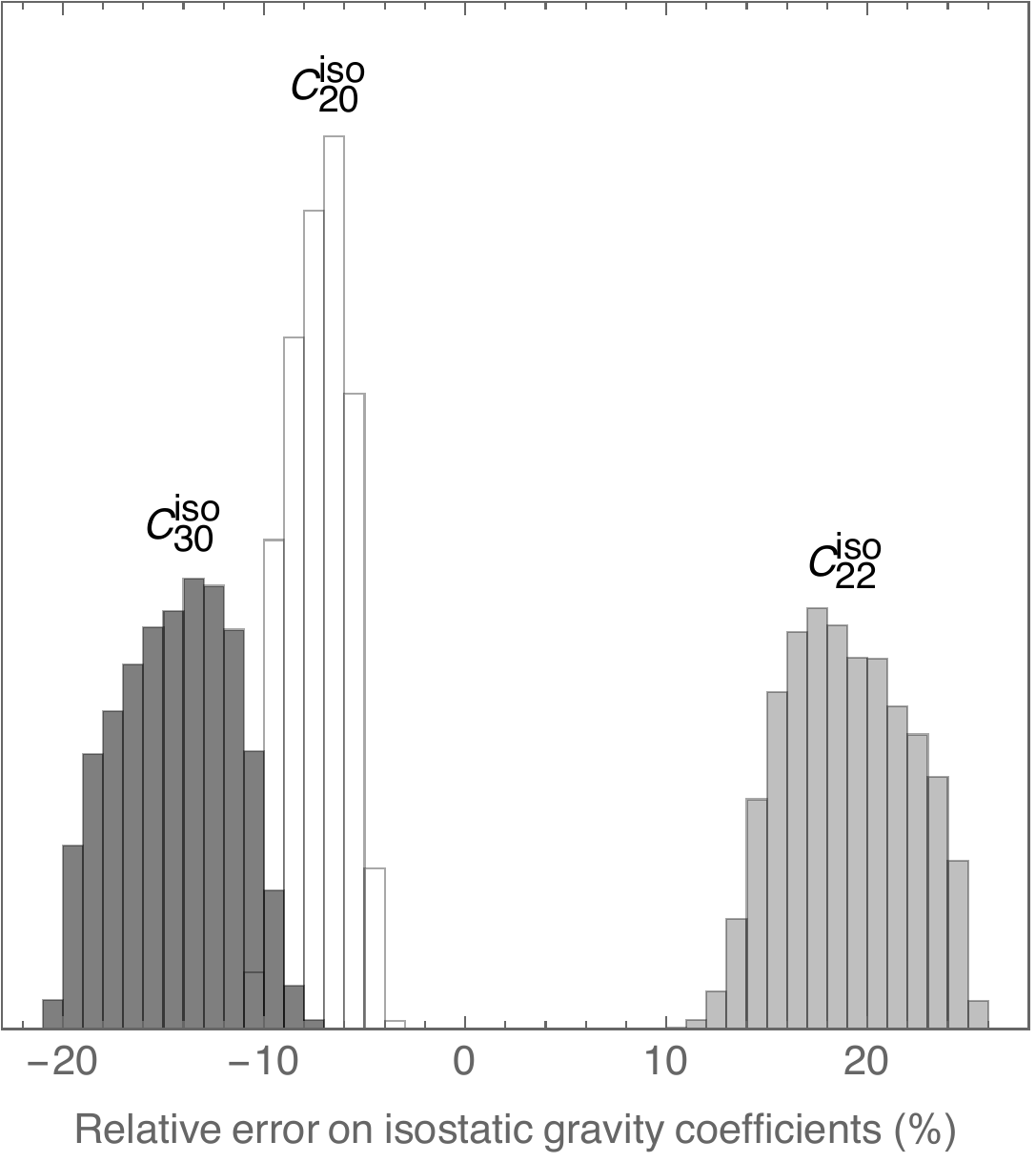}
\end{subfigure}%
\hspace{-1cm}
\begin{subfigure}[b]{.5\linewidth}
\captionsetup{skip=-0.cm}
\caption{}\hspace{3.9cm}\label{fig:errorb}
   \includegraphics[width=6cm]{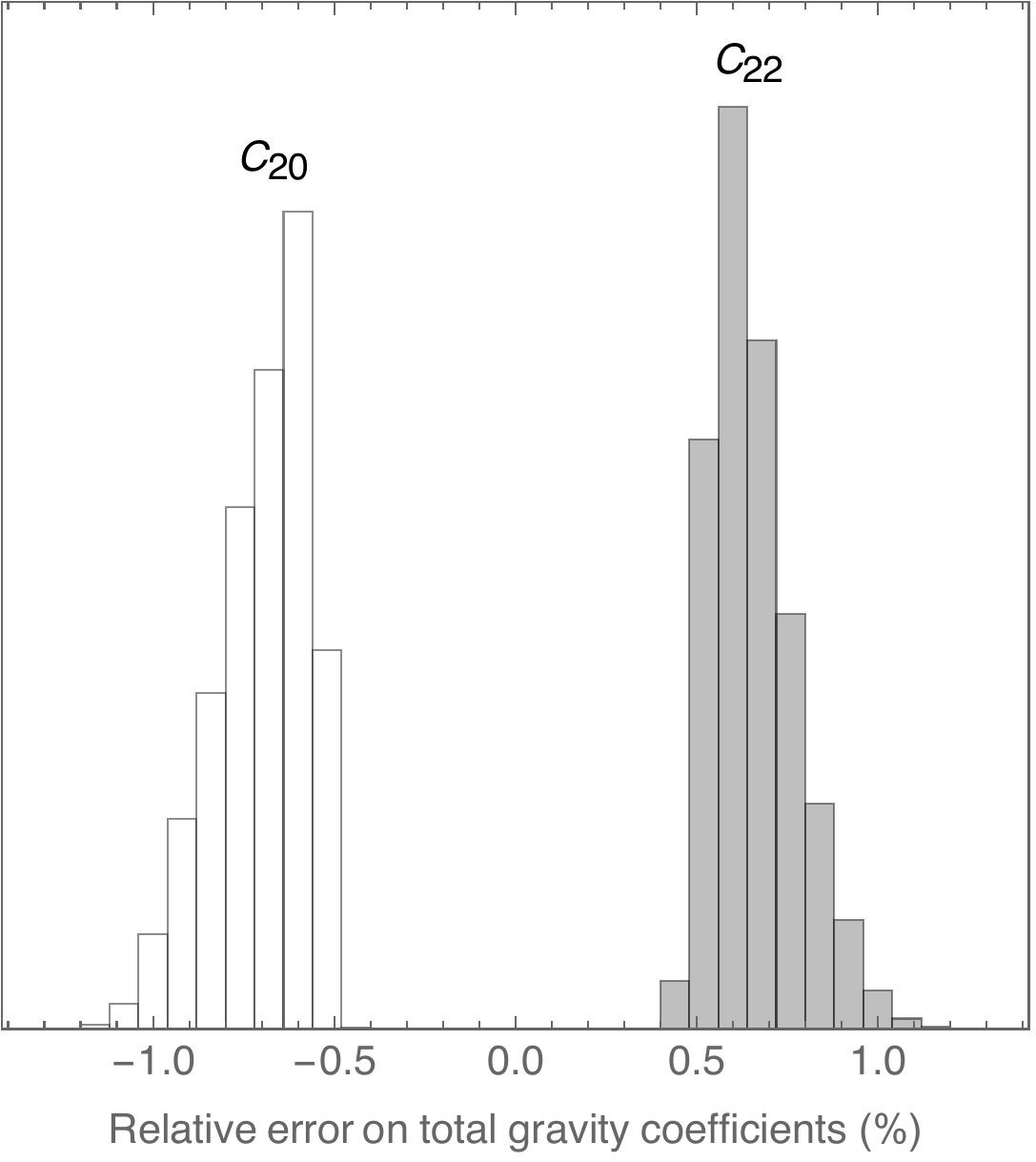}
\end{subfigure}%

   \caption[Second-order error on the isostatic model]{
   Second-order error on the isostatic model.
   Distribution of the relative error on (a) the isostatic gravity coefficients, (b) the total gravity coefficients.
The satellite is Enceladus and the model is SOL1/TOPA.}
   \label{FigError}
\end{figure}

\section{Second-order error on the isostatic model (Figure~\ref{FigError})}
\label{TextS6}

In this paper, gravity and shape are split into hydrostatic and isostatic components.
Second-order corrections are included in the figure of equilibrium but not in the isostatic model.
In order to estimate the error due to this approximation, we compute the second-order gravitational potential $C_{nm}^{tot}$ generated by the total shape (hydrostatic plus isostatic) of all density interfaces (core/ocean, ocean/crust, and surface).
The second-order error on the isostatic model is estimated by the difference $C_{nm}^{tot}-(C_{nm}^h+C_{nm}^{iso})$.

Consider first a spherically symmetric body of radius $R$ stratified into layers of uniform density.
Next, add topographic relief to the interface of radius $D$ between two layers which have a positive density contrast $\Delta\rho$.
Powers of the relief can be expanded in real spherical harmonics: $(h/D)^p=s^{(p)}_{inm}Y_{inm}(\theta,\phi)$, where the first subscript denotes cosine ($i=1$) or sine ($i=2$) components.
The gravitational potential caused by this relief at the surface is given by \citep{wieczorek1998}
\begin{equation}
U = 4\pi G D^2 \Delta\rho \sum_{i,n,m} \left(\frac{D}{R}\right)^{n+1} \frac{s^{\rm FA}_{inm}}{2n+1} \, Y_{inm}(\theta,\phi) \, ,
\end{equation}
where the finite amplitude (FA) contribution of the shape reads
\begin{equation}
s^{\rm FA}_{inm} = \sum_{p=1}^{n+3} \left( \frac{s^{(p)}_{inm}}{p!} \right) \left( \frac{\prod_{j=1}^p(n+4-j)}{n+3} \right) .
\end{equation}
If the only nonzero shape coefficients are $(s_{20},s_{22},s_{30})=(s^{(1)}_{120},s^{(1)}_{122},s^{(1)}_{130})$, the finite amplitude terms associated to these coefficients are given to second order ($p\leq2$) by
\begin{eqnarray}
s^{\rm FA}_{20} &=& s_{20} + \frac{4}{7} \left( (s_{20})^2 - 12 \, (s_{22})^2 + \frac{2}{3} \, (s_{30})^2 \right)  ,
\nonumber \\
s^{\rm FA}_{22} &=& s_{22} - \frac{8}{7} \, s_{20} \, s_{22} \, ,
\nonumber \\
s^{\rm FA}_{30} &=& s_{30} + \frac{4}{3} \, s_{20} \, s_{30} \, .
\end{eqnarray}

For Enceladus, the second-order error on the isostatic model is 5 to 10\% of $C_{20}^{iso}$, 10 to 26\% of $C_{22}^{iso}$, and 10 to 20\% of $C_{30}^{iso}$ (Fig.~\ref{fig:errora}).
However, the degree-two isostatic components are a small fraction of the total gravity coefficient (less than 10\% of $C_{20}$ and 3\% of $C_{22}$ for Enceladus).
Therefore the second-order error on the isostatic model is less than 1\% of the degree-two total gravity coefficients and less than 20\% of the degree-three gravity coefficient, which is comparable to the uncertainty on the data.
Finite amplitude corrections are smaller for Dione.

\section{Comment on a paper by \textit{Cadek et al. (2016)}}
\label{TextS7}

\par\noindent
{\bf Outline.}
A recent paper by \citet{cadek2016} claims that Enceladus's lithosphere provides flexural elastic support so that the shell can be thinner than predicted by classical isostasy.
We show here that their claim is wrong.

In that paper, the classical Airy relation between the surface and compensating topography is replaced by one depending on an additional parameter: the elastic thickness $T_e$ of the lithosphere (see Eq.~(\ref{cadekeq1}) below).
Because of this new free parameter, the gravity-shape problem becomes underdetermined (see their Fig.~2) and the average shell thickness must be constrained by independent information provided by librations.
{\it Cadek et al.} tune $T_e$ so that the shell thickness matches the one required in the libration model ($20\rm\,km$).
Thus it is not surprising that they draw the same conclusions about a thin south polar crust as do the libration models with nonhydrostatic crustal boundaries \citep{vanhoolst2016}.

We will make two points:
\begin{enumerate}
\item
The flexural-isostatic model used in \citet{cadek2016} is wrong: it is based on an equation for top loading, but the degree of compensation is computed with a formula valid for bottom loading.
If flexural isostasy results from top loading, $T_e$ should be about $20\rm\,m$ instead of $200\rm\,m$ as stated in the paper.
\item
Supporting more than $1\rm\,km$ of topography with a $20\rm\,m$-thick lithosphere stretches belief.
More generally, whatever the exact thickness of the lithosphere (as long as it is thin), the degree-two deflection required to provide elastic support induces stresses that lead to lithospheric failure.
\end{enumerate}
We round off the section by discussing the problems raised by isostasy in thin shell theory.\\

\par\noindent
{\bf Flexural-isostatic model.}
In our notation, the flexural-isostatic model of \citet{cadek2016} (their Eq.~(1)) reads
\begin{equation}
C_{el} \, \rho_s \, H_{nm}^{iso} \, g R^2 + \left(\rho_o-\rho_s\right) H_{nm}^{int} \, g_o R_o^2 = 0 \, ,
\label{cadekeq1}
\end{equation}
where $(R_o,g_o,H_{nm}^{int})$ are the radius, gravitational acceleration, and nonhydrostatic topography at the crust-ocean boundary, respectively.
The parameter $C_{el}$ is the degree of compensation for elastic flexure which varies between 0.75 and 0.9 in \citet{cadek2016} ($C_{el}=1$ in the limit of Airy isostasy).
With the approximation $g R^2/(g_o R_o^2)\approx1$, Eq.~(\ref{cadekeq1}) reduces to
\begin{equation}
H_{nm}^{int} = - \frac{\rho_s}{\rho_o-\rho_s} \, C_{el} \, H_{nm}^{iso} \, ,
\label{cadekeq2}
\end{equation}
where $H_{nm}^{int}$ should be interpreted as the deflection of the shell due to the surface topography $H_{nm}^{iso}$.
Eq.~(\ref{cadekeq2}) is the well-known compensation equation for top loading \citep{turcotte1981} (without geoid effects), and we expect a choice for $C_{el}$ that is consistent with top loading \citep{turcotte1981}:
\begin{equation}
C_{el} = C_{top} = \frac{\chi_n}{\tau \, \phi_n + \chi_n} \, ,
\label{Ctop}
\end{equation}
where  $\phi_n=n(n+1)-2$ and $\chi_n=\phi_n+1+\nu$ ($\nu$ is Poisson's ratio).
The bending resistance is neglected here so that there is only one parameter measuring the rigidity of the shell \citep{turcotte1981}:
\begin{equation}
\tau = \frac{ET_e}{R^2g(\rho_o-\rho_s)} \, ,
\label{tau}
\end{equation} 
where $E$ is Young's modulus and $T_e$ is the elastic thickness of the lithosphere.

Instead, {\it Cadek et al.}\  define $C_{el}$ by a formula which was proposed for the modeling of dynamic topography (Eq.~(25) of \citet{kalousova2012}), a kind of bottom loading:
\begin{equation}
C_{el} = C_{bot} = \frac{\chi_n}{\tau' \, \phi_n + \chi_n} \, ,
\label{Cbot}
\end{equation}
where $\tau'$ is defined by
\begin{equation}
\tau' = \frac{ET_e}{R^2g\rho_o} \, .
\label{tauprime}
\end{equation}
The correct formula for the shell deflection due to bottom loading would be (Eq.~(A10) of \citet{mcgovern2002}):
\begin{equation}
H_{nm}^{int} = - \frac{\rho_o}{\rho_o-\rho_s} \, \frac{1}{C_{bot}} \, H_{nm}^{iso} \, ,
\label{bottomload}
\end{equation}
where $H_{nm}^{int}$ should be interpreted as the thickness of the bottom load and $H_{nm}^{iso}$ as the shell deflection.
Clearly, there is a contradiction in using Eq.~(\ref{cadekeq2}) with $C_{el}$ defined by Eq.~(\ref{Cbot}).
The correct way to proceed would be to use Eq.~(\ref{cadekeq2}) with the degree of compensation for top loading, Eq.~(\ref{Ctop}).

\begin{figure}
   \centering
   \includegraphics[width=7cm]{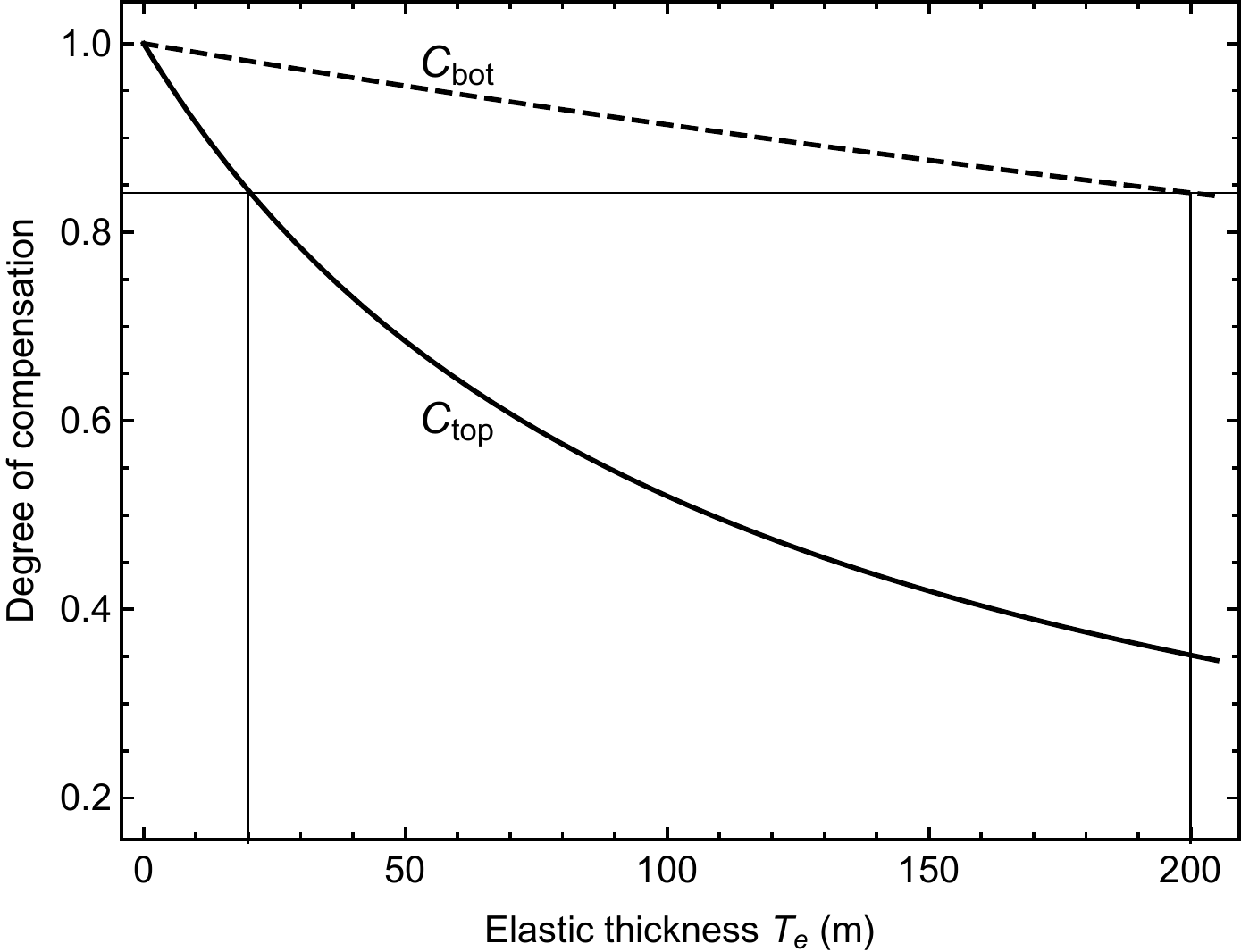}
   \caption[Degree of compensation for top and bottom loading in Enceladus]{
   Degree of compensation for top and bottom loading in Enceladus.
   The factors $C_{top}$ (top loading, solid curve) and $C_{bot}$ (bottom loading, dashed curve) are given by Eqs.~(\ref{Ctop}) and (\ref{Cbot}), respectively.
   The straight lines indicate the $T_e$ values corresponding to $C_{top}=0.84$ and $C_{bot}=0.84$.
   The densities are $(\rho_o,\rho_s)=(1030,925)\rm\,kg/m^3$ and the elastic parameters are $(E,\nu)=(9.3\rm\,GPa,0.33)$.
   }
   \label{FigCompens}
\end{figure}

If $\rho_o=1030\rm\,kg/m^3$ and $\rho_s=925\rm\,kg/m^3$, $\tau'\approx\tau/10$ and $C_{bot}$ is much larger than $C_{top}$ (Fig.~\ref{FigCompens}).
In particular, $C_{bot}=0.84$ corresponds to $T_e=200\rm\,m$, which is the elastic thickness favored in \citet{cadek2016}.
By contrast, using the correct formula for the degree of compensation ($C_{top}=0.84$) yields $T_e=20\rm\,m$.
It is hardly credible that more than $1\rm\,km$ of topography can stand on such a thin lithosphere without breaking it (see below).

Flexural isostasy with top loading is not the only possibility.
Alternatively, Eqs.~(\ref{Cbot}) and (\ref{bottomload}) can be used if the deflection of the shell (or crust) results from a subsurface density anomaly.
Beware that the thickness of the bottom load is not identical to the topography of the crust-ocean boundary: the former is measured with respect to the deflected crust-ocean boundary whereas the latter is measured with respect to the spherical surface coinciding with the crust-ocean boundary before deflection.
If top and bottom loads are both present, the flexure problem cannot be solved without specifying another free parameter, which is the ratio of subsurface to surface loading (Eq.~(B5) of \citet{mcgovern2002}).\\

\par\noindent
{\bf Lithospheric failure.}
If we put aside the question of how $C_{el}$ is related to $T_e$, the flexural-isostatic model of \citet{cadek2016} (Eq.~(\ref{cadekeq1})) can be seen as a parametric relation between surface topography and bottom topography, in which $C_{el}$ is a black-box parameter representing the effect of flexural isostasy.
Following \citet{mckinnon2013}, we now show that a thin lithosphere (less than $200\rm\,m$ thick) cannot support degree-two topography without breaking up, except in the limit of full isostasy with negligible deflection.

The degree-two zonal deformation of a spherical shell generates tangential membrane stresses $(\sigma_{\theta\theta},\sigma_{\phi\phi})$ given by the Vening-Meinesz equations (Eqs.~(4.2)-(4.3) of \citet{melosh2011}).
We quantify the stress in the lithosphere by the maximum differential stress
\begin{equation}
\Delta\sigma={\rm Max}(\sigma_{\theta\theta}-\sigma_{\phi\phi}) \approx \mu | \Delta f | \, ,
\end{equation}
where $\mu=3.5\rm\,GPa$ is the crustal shear modulus and $\Delta{}f$ is the shell flattening.
The latter is related to the $(2,0)$ harmonic component of the shell deflection ($H_{20}^{flex}$) by $\Delta{}f=(-3/2)H_{20}^{flex}/R$.
We neglect here the equatorial flattening because Enceladus's nonhydrostatic deformation is mostly zonal.

If compensation results from top loading, $H_{20}^{flex}$ is equal to the deflection of the shell given by $H_{20}^{int}$ in Eq.~(\ref{cadekeq1}), which is much larger in magnitude than the topographic load itself because of the near-complete compensation ($C_{el}=0.84$).
The hydrostatic-isostatic decomposition yields $H_{20}^{iso}\approx-1300\rm\,m$ so that the maximum differential stress is $\Delta\sigma\approx240\rm\,MPa$.
If compensation results from bottom loading, $H_{20}^{flex}$ is equal to the shell deflection ($H_{20}^{iso}$ in Eq.~(\ref{bottomload})), so that the maximum differential stress is $\Delta\sigma\approx30\rm\,MPa$.
Both estimates are much larger than the tensional strength of intact cold ice (1 to $2\rm\,MPa$) \citep{schulson2009}.
The lithosphere actually fails at lower stress levels because of its finite brittle strength, as discussed in detail in \citet{mckinnon2013}.
In conclusion, the nonhydrostatic degree-two topography on Enceladus cannot be supported by lithospheric flexure.
The only way out consists in distributing the loads on the top and bottom of the crust so as to minimize the shell deflection and the resulting membrane stresses.
But this state of full isostasy does not depend on the elastic thickness of the lithosphere.\\

\par\noindent
{\bf Isostasy in thin shell theory.}
Applying thin shell theory to isostasy is problematic for three reasons.
First, thin shell theory is not valid if the shell is thicker than ten percent of the shell radius.
Thus one must suppose a priori that the ice shells of Enceladus and Dione are less than $25\rm\,km$ and $56\rm\,km$ thick, respectively.
Enceladus's ice shell seems to satisfy, but just barely, the thin shell threshold whereas Dione's ice shell does not.
Thin shell theory could be valid for Dione (or Enceladus) if the lithosphere is much thinner than the whole ice shell.
It is not obvious, however, how the subsurface loading applied on the crust-ocean boundary is transmitted to the lithosphere itself (see third reason below).
Second, the geoid cannot be included in a consistent way: the topographic loads are measured with respect to the geoid which is affected, in turn, by the deflection of the shell as well as by the surface and subsurface loads themselves.
This effect is crucial at long wavelengths.
Third, the standard (first-order) formalism of thin shell theory does not take into account the difference in area between the inner and outer shell surfaces.
It is thus not possible to introduce ad hoc factors $g R^2$ and $g_o R_o^2$ as was done by \textit{Cadek et al.}\ (compare Eqs.~(\ref{cadekeq1}) and (\ref{cadekeq2})), since factors of similar magnitude are neglected elsewhere in thin shell theory.
This problem must be solved by working with a higher-order approximation of thin shell theory which is not worth the effort in comparison with the minimum stress isostasy developed in our paper.

\clearpage


\clearpage


\begin{table}[h]\centering
\small
\caption[General parameters for Enceladus and Dione]{
General parameters for Enceladus and Dione.
Surface radii are taken from Tables~\ref{TableTopoEnc} and \ref{TableTopoDione}.
Gravitational parameters are taken from SOL1 of \citet{iess2014} for Enceladus and from JPL online data (http://ssd.jpl.nasa.gov/) for Dione.
The bulk density and surface gravity are computed from $R$ and $GM$ (Tables~\ref{TableGraviEnc}--\ref{TableGraviDione}).
The rotational parameter is defined by $q=\omega^2R^3/GM$ with $\omega=2\pi/T$.}
\begin{tabular}{@{}lrrrr@{}}
\hline
Parameter &  Symbol & Enceladus  &  Dione & Unit \\
\hline
Surface radius & $R$  & 252.1& 561.4 & km \\
Gravitational parameter & $GM$ & $7.21044$ & $73.1146$ & $\rm{}km^3/s^2$ \\
Bulk density & $\bar\rho$  & 1610 & 1478 & $\rm{}kg/m^3$ \\
Surface gravity & $g$ & 0.113 & 0.232 & $\rm{}m\,s^{-2}$ \\
Period & $T$ & 1.37 & $2.737$ & days \\
Rotational parameter & $q$  & $6.26\times10^{-3}$ & $1.71\times10^{-3}$ & - \\
\hline
\end{tabular}
\label{TableGeneral}
\end{table}%

\begin{table}[h]\centering
\small
\caption[Enceladus's gravity]{
Enceladus's gravity.
The nondimensional gravity coefficients $C_{nm}$ \citep{iess2014} are rescaled from the gravity field reference radius ($254.2\rm\,km$) to $R=252.1\rm\,km$ as in \citet{mckinnon2015}.
Error bars are $1\sigma$. $\varepsilon_1$ and $\varepsilon_2$ are the relative errors (in \%) for SOL1 and SOL2, respectively.
 See Text~\ref{TextS2}.}
\begin{tabular}{@{}lrrrr@{}}
\hline
Coeff. &  SOL1  &  $\varepsilon_1$ & SOL2  & $\varepsilon_2$ \\
& $(\times 10^6)$ & & $(\times 10^6)$ & \\
\hline
$C_{20}$ & $-5526\pm35$  & $0.6$ & $-5534\pm70$ & $1$ \\
$C_{22}$ & $1576\pm16$  & $1$ & $1640\pm53$ & $3$ \\
$C_{30}$ & $118\pm23$ & $20$ & $88\pm72$ & $80$ \\
\hline
$G_2$ & $3.51\pm0.04$  & $1$ & $3.37\pm0.12$ & $4$ \\
\hline
\end{tabular}
\label{TableGraviEnc}
\end{table}%

\begin{table}[h]\centering
\small
\caption[Dione's gravity]{
Dione's gravity.
SOL1 is the preliminary solution given in \citet{hemingway2016}.
Same notation as in Table~\ref{TableGraviEnc}.
}
\begin{tabular}{@{}lrr@{}}
\hline
Coeff. &  SOL1  &  $\varepsilon_1$ \\
& $(\times 10^6)$ \\
\hline
$C_{20}$ & $-1454\pm16$  & $1$ \\
$C_{22}$ & $363\pm2$  & $0.6$  \\
\hline
$G_2$ & $4.01\pm0.05$  & $1$ \\
\hline
\end{tabular}
\label{TableGraviDione}
\end{table}%

\begin{table}[h]\centering
\small
\caption[Enceladus's shape]{
Enceladus's shape.
TOPA and TOPB are the shape solutions of \citet{nimmo2011} and \citet{thomas2016}, respectively, expressed in terms of unnormalized real spherical coefficients $H_{nm}$.
$R_0$ is the radius of the sphere of equal volume.
As in the original papers, error bars are $2\sigma$ for $(a,b,c,R_0,F)$ and $1\sigma$ for $(H_{nm},F_2)$.
$\varepsilon_A$ and $\varepsilon_B$ are the relative errors (in \%) on the harmonic coefficients of TOPA and TOPB, respectively.
GEO1 and GEO2 are the geoids computed to first order with the gravity solutions SOL1 and SOL2, respectively.
See Text~S3.}
\begin{tabular}{@{}rrrrrrrr@{}}
\hline
& TOPA & $\epsilon_A$ & TOPB  & $\epsilon_B$ & GEO1 & GEO2 & Unit \\
\hline
$a$ & $256.8\pm0.2^*$ & & $256.2\pm0.3$ & & & & km \\
$b$ & $251.3\pm0.2^*$ & & $251.4\pm0.2$ & & & & km \\
$c$ & $248.3\pm0.4^*$ & & $248.6\pm0.2$ & & & & km \\
$R_0$ & $252.1\pm0.2$ & & $252.0\pm0.2$ & & & & km \\
\hline
$H_{20}$ & $-3846\pm179$ & 5 & $-3470\pm90^{**}$ & 3 & $-2708\pm9$ & $-2710\pm18$ & m \\
$H_{22}$ & $917\pm19$ & 2 & $800\pm30^{**}$ & 4 & $792\pm4$ & $808\pm13$ & m \\
$H_{30}$ & $384\pm5$ & 1 & -  & - & $30\pm6$ & $22\pm18$ & m \\
$F_2$ & $4.20\pm0.21$ & 5 & $4.33\pm0.20$ & 5 & - & - & - \\
\hline
\multicolumn{8}{l}{ ${}^*$ Error bars on the differences $(a-R_0,b-R_0,c-R_0)$.}\\
\multicolumn{8}{l}{ ${}^{**}$ Error bars assuming that errors on $(a,b,c)$ add quadratically.}
\end{tabular}
\label{TableTopoEnc}
\end{table}%

\begin{table}[h]\centering
\small
\caption[Dione's shape]{
Dione's shape.
See Table~\ref{TableTopoEnc} for details.}
\begin{tabular}{@{}rrrrr@{}}
\hline
& TOPA & $\epsilon_A$ & GEO1 & Unit \\
\hline
$a$ & $563.5\pm0.3^*$ & & & km \\
$b$ & $561.3\pm0.3^*$ & & & km \\
$c$ & $559.5\pm0.3^*$ & & &  km \\
$R_0$ & $561.4\pm0.3$ & & & km \\
\hline
$H_{20}$ & $-1923\pm157$ & 8 & $-1616\pm9$ & m \\
$H_{22}$ & $368\pm32$  & 9 & $444\pm1$ &  m \\
$F_2$ & $5.2\pm0.6$ & 12 & - & - \\
\hline
\multicolumn{5}{l}{ ${}^*$ Error bars on $(a-R_0,b-R_0,c-R_0)$.}
\end{tabular}
\label{TableTopoDione}
\end{table}%

\clearpage

\begin{table}[h]\centering
\small
\caption[Enceladus: mean values and $1\sigma$ (Bayesian) confidence intervals]{
Enceladus: mean values and $1\sigma$ (Bayesian) confidence intervals.
The inverted parameters are the shell thickness $d_s$, ocean thickness $d_o$, shell density $d_s$, and ocean density $d_o$.
The core radius $r_c$ and core density $\rho_c$ are derived parameters without prior.
The prior on $\rho_s$ is not valid for the POROUS model (see Section~3.4).
The symbol $*$ indicates that the parameter is not constrained by the inversion.
The numbers $(nm,...)$ in the second row denote the degree $n$ and order $m$ of the harmonic coefficients used in the inversion.
The layer thicknesses sum to $R\pm1\rm\,km$ because the results are rounded to the nearest integer.
 See Text~\ref{TextS5}.}
\begin{tabular}{@{}lcrrrrr@{}}
\hline
Parameter &  Prior range & SOL1/TOPA & SOL2/TOPA &  SOL1/TOPA & SOL1/TOPB & POROUS \\
& & $(20,22,30)$ & $(20,22,30)$ & $(20,22)$ & $(20,22)$ & $(20,22,30)$ \\
\hline
$d_s$ (km) & $1-60$ &  $23\pm4$ & $24\pm7$ & $31\pm9$ & $31\pm9$ &  $27\pm4$  \\
$d_o$ (km) & $5-60$ & $38\pm4$ & $34\pm6$ & $31\pm8$ & $31\pm8$ &  $29\pm5$ \\
$r_c$ (km) & $-$ & $192\pm2$ & $194\pm4$ & $190\pm3$ & $190\pm3$ &  $196\pm3$ \\
$\rho_s$ ($\rm{}kg/m^3$) & $880-960$ & $*$ & $*$ & $*$ & $*$ &  $*$  \\
$\rho_o$ ($\rm{}kg/m^3$) & $1000-1040$ & $*$ & $*$ & $*$ & $*$ &  $*$  \\
$\rho_c$ ($\rm{}kg/m^3$) & $-$ & $2422\pm46$ & $2377\pm94$ & $2477\pm76$ & $2479\pm76$ &  $2370\pm44$ \\
\hline
\end{tabular}
\label{TableInvEnc}
\end{table}%

\begin{table}[h]\centering
\small
\caption[Dione: mean values and $1\sigma$ (Bayesian) confidence intervals]{
Dione: mean values and $1\sigma$ (Bayesian) confidence intervals.
See Table~\ref{TableInvEnc} for the notation.
TOPB is the same as TOPA but with zero error bars.
The layer thicknesses sum to $562\rm\,km$ because the results are rounded to the nearest integer.
 See Text~\ref{TextS5}.}
\begin{tabular}{@{}lcrrr@{}}
\hline
Parameter & Prior range & SOL1/TOPA & SOL1/TOPB \\
\hline
$d_s$ (km) & $1-200$ & $99\pm23$ & $102\pm10$  \\
$d_o$ (km) & $5-200$ & $65\pm30$ & $57\pm16$ \\
$r_c$ (km) & $-$ & $398\pm14$ & $403\pm9$ \\
$\rho_s$ ($\rm{}kg/m^3$) & $880-960$ & $*$ & $*$ \\
$\rho_o$ ($\rm{}kg/m^3$) & $1000-1040$ & $*$ & $*$ \\
$\rho_c$ ($\rm{}kg/m^3$) & $-$ & $2435\pm140$ &$2383\pm58$ \\
\hline
\end{tabular}
\label{TableInvDione}
\end{table}%

\end{document}